\documentclass[prd,aps,preprint,showpacs,nofootinbib,superscriptaddress]{revtex4}
\usepackage[latin1]{inputenc}
\usepackage{epsfig}
\usepackage{amsmath}
\usepackage{latexsym}
\usepackage{amssymb}
\usepackage{dsfont}
\usepackage{color}
\usepackage{multirow}

\newcommand{\ud}{\mathrm{d}}

\newlength\savedwidth
\newcommand\whline{\noalign{\global\savedwidth\arrayrulewidth
\global\arrayrulewidth 1pt}%
\hline
\noalign{\global\arrayrulewidth\savedwidth}}

\begin{document}


\title{Quark Wigner Distributions and  Orbital Angular Momentum}
\author{C. Lorc\'e\footnote{E-mail: lorce@kph.uni-mainz.de}}
\affiliation{Institut f\"ur Kernphysik, Johannes Gutenberg-Universit\"at,\\ D-55099 Mainz, Germany}

\author{B. Pasquini\footnote{E-mail: pasquini@pv.infn.it}}
\affiliation{Dipartimento di Fisica Nucleare e Teorica, Universit\`a degli Studi di Pavia, \\and INFN, Sezione di Pavia, I-27100 Pavia, Italy}

\begin{abstract}
We study the  Wigner functions of the nucleon  which provide multidimensional images of the quark distributions in phase space. These functions can be obtained through a Fourier transform in the transverse space of the generalized transverse-momentum dependent parton distributions. They depend on both the transverse position and the three-momentum of the quark relative to the nucleon, and therefore combine in a single picture all the information contained in the generalized parton distributions and the transverse-momentum dependent parton distributions. We focus the discussion on the distributions of unpolarized/longitudinally polarized quark in an unpolarized/longitudinally polarized nucleon. In this way, we can study  the role of the orbital angular momentum of the quark in shaping the nucleon and its correlations with the quark and nucleon polarizations. The quark orbital angular momentum is also calculated from its phase-space average weighted with the Wigner distribution of unpolarized quarks in a longitudinally polarized nucleon. The corresponding results obtained within different light-cone quark models are compared with alternative definitions of the quark orbital angular momentum, as given in terms of generalized parton distributions and transverse-momentum dependent parton distributions. 
\end{abstract}


\maketitle

\section{Introduction}
\label{section-1}

One of the most challenging tasks for unravelling the partonic structure of hadrons is mapping the distribution of momentum and spin of the proton onto its constituents. To this aim, generalized parton distributions (GPDs) \cite{Mueller:1998fv,Ji:2004gf,Belitsky:2005qn,Goeke:2001tz,Diehl:2003ny,Boffi:2007yc} and transverse-momentum dependent parton distributions (TMDs) \cite{Collins:1981uk,Collins:1981uw,Sivers:1989cc,Kotzinian:1994dv,Mulders:1995dh,Boer:1997nt} have proven to be among the most useful tools. GPDs provide a new method of spatial imaging of the nucleon \cite{Burkardt:2005td,Burkardt:2002hr,Burkardt:2000za,Diehl:2005jf,Pasquini:2007xz}, through the definition of impact-parameter dependent densities (IPDs) which reveal the correlations between the quark distributions in transverse-coordinate (or impact-parameter) space and longitudinal momentum for different quark and target polarizations. On the other hand, TMDs contain novel and direct three-dimensional information about the strength of different spin-spin and spin-orbit correlations in the momentum space \cite{Goeke:2005hb,Bacchetta:2006tn,Meissner:2007rx,Lorce:2011zt}. The ultimate understanding  of the partonic structure of the nucleon can be gained by means of joint position-and-momentum (or phase-space) distributions such as the Wigner distributions. These distributions contain the most general one-body information of partons, corresponding to the full one-body density matrix in both momentum and position space, and reduce in certain limits to TMDs and GPDs. Because of the uncertainty principle which prevents to know simultaneously the position and momentum of a quantum-mechanical system, the phase-space distributions do not have a density interpretation. Only in the classical limit they become positive definite. Nonetheless, the physics of a phase-space distribution is very rich and one can try to select certain situations where a semi-classical interpretation is still possible. Wigner distributions have already been applied in many physics areas like heavy ion collisions, quantum molecular dynamics, signal analysis, quantum information, optics, image processing, nonlinear dynamics, \ldots ~\cite{Balazs:1983hk,Hillery:1983ms,Lee:1995}, and can even be measured directly in some experiments \cite{Vogel:1989zz,Smithey:1993zz,Breitenbach:1997,Banaszek:1999ya}.

The concept of Wigner distributions in QCD for quarks and gluons was first explored in Refs.~\cite{Ji:2003ak,Belitsky:2003nz}, introducing the definition of a Wigner operator whose matrix elements in the nucleon states were interpreted as distributions of the partons in a six-dimensional space (three position and three momentum coordinates). The link with GPDs was exploited to obtain three-dimensional spatial images of the proton which were interpreted as charge distributions of the quarks for fixed values of the Feynman variable $x$. This interpretation relies however on a nonrelativistic approximation.

Wigner distributions have a direct connection with the generalized parton correlation functions (GPCFs) which were recently introduced in Ref.~\cite{Meissner:2009ww}. The GPCFs are the distributions that parametrize the fully unintegrated, off-diagonal quark-quark correlator for a hadron. In the case of the nucleon and after integration over the light-cone energy of the quark, one finds the so-called generalized transverse-momentum dependent parton distributions (GTMDs). At leading-twist there are 16 GTMDs which depend on the light-cone three-momentum of the quark and, in addition, on the momentum transfer to the nucleon $\Delta^\mu$. After two-dimensional Fourier transform from $\vec \Delta_\perp$ to the impact-parameter space coordinates $\vec b_\perp$, in a frame without momentum transfer along the light-cone direction, one obtains the Wigner distributions which are completely consistent with special relativity.

The purpose of this paper is to investigate the phenomenology of the quark Wigner distributions. As a matter of fact, since it is not known how to access these distributions directly from experiments, phenomenological models are very powerful in this context. Collecting the information that one can learn from quark models which were built up on the basis of available experimental information on GPDs and TMDs, one can hope to reconstruct a faithful description of the physics of the Wigner distributions. To this aim we will rely on models for the light-cone wave functions (LCWFs) which have already been used for the description of the generalized parton distributions (GPDs)~\cite{Boffi:2002yy,Pasquini:2005dk,Boffi:2007yc,Lorce:2011dv}, the transverse-momentum dependent parton distributions (TMDs)~\cite{Pasquini:2008ax,Boffi:2009sh,Pasquini:2010af,Lorce:2011dv,Lorce:2011zt} and  electroweak properties of the nucleon~\cite{Lorce:2006nq,Lorce:2007as,Lorce:2007fa,Pasquini:2007iz,Lorce:2011dv}.

The plan of the manuscript is as follows. In sec.~\ref{section-2a}, we review the definition of the Wigner distributions obtained by Fourier transform of the GTMDs to the impact-parameter space. Although the Wigner distributions cannot have a strict probabilistic interpretation, they reduce to genuine probability distributions after integration over position and/or momentum variables. As discussed in sec.~\ref{section-2b}, one can obtain four types of three dimensional densities: in addition to momentum distributions given by the TMDs and to IPDs related to the GPDs, there are two new distributions mapping the nucleon as functions of one transverse-space coordinate and one transverse-momentum component which are not conjugated and therefore not constrained by the Heisenberg uncertainty principle. Whereas the GTMDs are in general complex-valued functions, the two-dimensional Fourier transforms of the GTMDs are always real-valued functions, in accordance with their interpretation as phase-space distributions. These 16 functions can be disentangled by selecting different configurations, along three orthogonal directions, of the nucleon and quark polarizations. In order to simplify the discussion, in sec.~\ref{section-2c} we focus on the cases without transverse polarizations. In sec.~\ref{section-2d} we discuss and compare different definitions of the quark orbital angular momentum, as obtained from GPDs, TMDs and Wigner distributions. In particular, by treating the Wigner functions as if they were classical distributions, we can obtain the expectation value of the orbital angular momentum operator from its phase-space average weighted with the Wigner distribution of unpolarized quarks in a longitudinally polarized nucleon. In sec.~\ref{section-3} we explicitly calculate the Wigner distributions in two light-cone quark models, showing the results for the first $x$ moments in the four-dimensional phase space (two transverse position and two transverse momentum coordinates). In particular, we discuss specific situations where the density matrices have a quasi-probabilistic interpretation, giving a semi-classical picture for multidimensional images of the nucleon. In section~\ref{section-4} we draw our conclusions.

\section{Wigner Distributions}
\label{section-2}

\subsection{Wigner Operators and Wigner Distributions}
\label{section-2a}

Wigner distributions in QCD were first explored in Refs.~\cite{Ji:2003ak,Belitsky:2003nz}. Neglecting relativistic effects, the authors used the standard three-dimensional Fourier transform in the Breit frame and introduced six-dimensional Wigner distributions (three position and three momentum coordinates). We propose to study instead five-dimensional Wigner distributions (two position and three momentum coordinates) as seen from the infinite momentum frame (IMF). The advantages of working in the IMF have already been emphasized in the derivation of transverse charge densities~\cite{Miller:2007uy,Miller:2010nz,Carlson:2007xd} and IPDs~\cite{Burkardt:2000za,Burkardt:2002hr,Burkardt:2005td,Diehl:2005jf} from form factors (FFs) and GPDs, respectively. Analogously, they will be exploited  here to arrive at a definition of Wigner distributions which is not spoiled by relativistic corrections.

Introducing two lightlike four-vectors $n_\pm$ satisfying $n_+\cdot n_-=1$, we write the light-cone components of a generic four-vector $a$ as $\left[a^+,a^-,\vec a_\perp\right]$ with $a^\pm=a\cdot n_\mp$. Similarly to Refs.~\cite{Ji:2003ak,Belitsky:2003nz}, we define the Wigner operators for quarks at a fixed light-cone time $y^+=0$ as follows
\begin{equation}\label{wigner-operator}
\widehat W^{[\Gamma]}(\vec b_\perp,\vec k_\perp,x)\equiv\frac{1}{2}\int\frac{\ud z^-\,\ud^2z_\perp}{(2\pi)^3}\,e^{i(xp^+z^--\vec k_\perp\cdot\vec z_\perp)}\,\overline{\psi}(y-\tfrac{z}{2})\Gamma\mathcal W\,\psi(y+\tfrac{z}{2})\big|_{z^+=0}
\end{equation}
with $y^\mu=[0,0,\vec b_\perp]$, $p^+$ the average nucleon longitudinal momentum and $x=k^+/p^+$ the average fraction of nucleon longitudinal momentum carried by the active quark. The superscript $\Gamma$ stands for any twist-two Dirac operator $\Gamma=\gamma^+,\gamma^+\gamma_5,i\sigma^{j+}\gamma_5$ with $j=1,2$. A Wilson line $\mathcal W\equiv\mathcal W(y-\tfrac{z}{2},y+\tfrac{z}{2}|n)$ ensures the color gauge invariance of the Wigner operator, connecting the points $(y-\tfrac{z}{2})$ and $(y+\tfrac{z}{2})$ \emph{via} the intermediary points $(y-\tfrac{z}{2})+\infty\cdot n$ and $(y+\tfrac{z}{2})+\infty\cdot n$ by straight lines \cite{Meissner:2009ww}. 

We define the Wigner distributions in terms of the matrix elements of the Wigner operators sandwiched between nucleon states with polarization $\vec S$ as follows
\begin{equation}\label{wigner}
\rho^{[\Gamma]}(\vec b_\perp,\vec k_\perp,x,\vec S)\equiv\int\frac{\ud^2\Delta_\perp}{(2\pi)^2}\,\langle p^+,\tfrac{\vec\Delta_\perp}{2},\vec S|\widehat W^{[\Gamma]}(\vec b_\perp,\vec k_\perp,x)|p^+,-\tfrac{\vec\Delta_\perp}{2},\vec S\rangle.
\end{equation}
Thanks to the properties of the Galilean subgroup of transverse boosts in the IMF \cite{Burkardt:2005td,Kogut:1969xa}, we can form a localized nucleon state in the transverse direction, in the sense that its transverse center of momentum is at the position $\vec r_\perp$ :
\begin{equation}
|p^+,\vec r_\perp\rangle=\int\frac{\ud^2p_\perp}{(2\pi)^2}\,e^{-i\vec p_\perp\cdot\vec r_\perp}|p^+,\vec p_\perp\rangle.
\end{equation}
The Wigner distributions defined according to Eq.~\eqref{wigner} can then be written in terms of these localized nucleon states as
\begin{equation}\label{wignerloc}
\rho^{[\Gamma]}(\vec b_\perp,\vec k_\perp,x,\vec S)=\int\ud^2D_\perp\,\langle p^+,-\tfrac{\vec D_\perp}{2},\vec S|\widehat W^{[\Gamma]}(\vec b_\perp,\vec k_\perp,x)|p^+,\tfrac{\vec D_\perp}{2},\vec S\rangle,
\end{equation}
where $\vec D_\perp$ is the transverse distance between the initial and final centers of momentum. Note that the nucleon in our definition of the Wigner distributions has vanishing average transverse position and average transverse momentum, see Eqs.~\eqref{wigner} and \eqref{wignerloc}. This allows us to interpret the variables $\vec b_\perp$ and $\vec k_\perp$ as the relative average transverse position and the relative average transverse momentum of the quark, respectively. 

Using a transverse translation in Eq.~\eqref{wigner}, we find
\begin{equation}
\rho^{[\Gamma]}(\vec b_\perp,\vec k_\perp,x,\vec S)=\int\frac{\ud^2\Delta_\perp}{(2\pi)^2}\,e^{-i\vec\Delta_\perp\cdot\vec b_\perp}\,W^{[\Gamma]}(\vec\Delta_\perp,\vec k_\perp,x,\vec S),
\end{equation}
where $W^{[\Gamma]}$ are the quark-quark correlators defining the GTMDs \cite{Meissner:2009ww} for $\Delta^+=0$
\begin{equation}
\begin{split}
W^{[\Gamma]}&(\vec\Delta_\perp,\vec k_\perp,x,\vec S)=\langle p^+,\tfrac{\vec\Delta_\perp}{2},\vec S|\widehat W^{[\Gamma]}(\vec 0_\perp,\vec k_\perp,x)|p^+,-\tfrac{\vec\Delta_\perp}{2},\vec S\rangle\\
&=\frac{1}{2}\int\frac{\ud z^-\,\ud^2z_\perp}{(2\pi)^3}\,e^{i(xp^+z^--\vec k_\perp\cdot\vec z_\perp)}\,\langle p^+,\tfrac{\vec\Delta_\perp}{2},\vec S|\overline{\psi}(-\tfrac{z}{2})\Gamma\mathcal W\,\psi(\tfrac{z}{2})|p^+,-\tfrac{\vec\Delta_\perp}{2},\vec S\rangle\big|_{z^+=0}.
\end{split}
\end{equation}
This means that  the Wigner distributions defined as in Eq.~\eqref{wigner} are the two-dimensional Fourier transforms of GTMDs, just like transverse densities and IPDs are two-dimensional Fourier transforms of FFs and GPDs, respectively. Contrarily to all the other distribution functions, the GTMDs are in general complex-valued functions. However the two-dimensional Fourier transforms of the GTMDs are always real-valued functions, in accordance with their interpretation as phase-space distributions.

\subsection{Three-Dimensional Probability Densities}
\label{section-2b}

Wigner distributions cannot have a strict probabilistic interpretation, since Heisenberg uncertainty relations prevent to  determine at the same time position and momentum of a particle. Accordingly, Wigner distributions are not positive definite. Nevertheless, integrating out position and/or momentum variables, Wigner distributions reduce to genuine probability distributions. There are in particular four types of three-dimensional probability densities:
\begin{itemize}
\item Integrating over $\vec b_\perp$ amounts to set $\vec\Delta_\perp=\vec 0_\perp$, and so the Wigner distributions reduce to the standard TMD correlators $\Phi^{[\Gamma]}$~\cite{Meissner:2009ww,Lorce:2011dv}
\begin{equation}
\begin{split}
\int\ud^2b_\perp\,\rho^{[\Gamma]}(\vec b_\perp,\vec k_\perp,x,\vec S)&=W^{[\Gamma]}(\vec 0_\perp,\vec k_\perp,x,\vec S)\\
&\equiv\Phi^{[\Gamma]}(\vec k_\perp,x,\vec S),
\end{split}
\end{equation}
which can be interpreted as quark densities in three-dimensional momentum space;
\item Integrating over $\vec k_\perp$ amounts to set $\vec z_\perp=\vec 0_\perp$, and so the Wigner distributions reduce to two-dimensional Fourier transforms of the standard GPD correlators \cite{Meissner:2009ww,Lorce:2011dv}
\begin{equation}
\int\ud^2k_\perp\,\rho^{[\Gamma]}(\vec b_\perp,\vec k_\perp,x,\vec S)=\int\frac{\ud^2\Delta_\perp}{(2\pi)^2}\,e^{-i\vec\Delta_\perp\cdot\vec b_\perp}\,F^{[\Gamma]}(\vec\Delta_\perp,x,\vec S)
\end{equation}
with
\begin{equation}
F^{[\Gamma]}(\vec\Delta_\perp,x,\vec S)\equiv\frac{1}{2}\int\frac{\ud z^-}{2\pi}\,e^{ixp^+z^-}\,\langle p^+,\tfrac{\vec\Delta_\perp}{2},\vec S|\overline{\psi}(-\tfrac{z}{2})\Gamma\mathcal W\,\psi(\tfrac{z}{2})|p^+,-\tfrac{\vec\Delta_\perp}{2},\vec S\rangle\big|_{z^+=z_\perp=0},
\end{equation}
where the modulus of a general transverse vector $\vec a_\perp$ is indicated as  $a_\perp$. In other words, one recovers the IPDs which can be interpreted as quark densities in the transverse position space and longitudinal momentum space;
\item Integrating over $b_y$ and $k_x$ amounts to set $\Delta_y=z_x=0$, and so the Wigner distributions reduce to new three-dimensional quark densities
\begin{equation}\label{bxky}
\int\ud b_y\,\ud k_x\,\rho^{[\Gamma]}(\vec b_\perp,\vec k_\perp,x,\vec S)\equiv\tilde\rho^{[\Gamma]}(b_x,k_y,x,\vec S).
\end{equation}
The variables $b_x$ and $k_y$ refer to two orthogonal directions in the transverse plane and so are not subjected to Heisenberg uncertainty relations;
\item Integrating over $b_x$ and $k_y$ amounts to set $\Delta_x=z_y=0$, and so the Wigner distributions reduce to other new three-dimensional quark densities
\begin{equation}\label{bykx}
\int\ud b_x\,\ud k_y\,\rho^{[\Gamma]}(\vec b_\perp,\vec k_\perp,x,\vec S)\equiv\bar\rho^{[\Gamma]}(b_y,k_x,x,\vec S).
\end{equation}
There are \emph{a priori} no simple relations between the quark densities in Eqs.~\eqref{bxky} and \eqref{bykx}, except when the quark and nucleon polarizations have no transverse components. In this case, the only privileged directions in the transverse plane are $\vec b_\perp$ and $\vec k_\perp$, and we have $\tilde\rho^{[\Gamma]}(b_\perp,k_\perp,x,\vec e_z)=\bar\rho^{[\Gamma]}(-b_\perp,k_\perp,x,\vec e_z)$ for $\Gamma=\gamma^+,\gamma^+\gamma_5$.
\end{itemize}

\subsection{Wigner Distribution with Longitudinal Polarizations}
\label{section-2c}

On the one hand, there are in total 16 GTMDs at twist-two level \cite{Meissner:2009ww}. On the other hand, the quark and nucleon can be either unpolarized or polarized along three orthogonal directions, which means 16 configurations. All the 16 configurations can be written in terms of 16 independent linear combinations of the GTMDs. We will not present all of them in this study. To keep the discussion relatively simple, we focus on cases without any transverse polarization.

The Wigner distribution of quarks with longitudinal polarization $\lambda$ in a nucleon with longitudinal polarization $\Lambda$ is obtained for $\Gamma=\gamma^+\tfrac{\mathds{1}+\lambda\gamma_5}{2}$ and $\vec S=\Lambda\vec e_z$
\begin{equation}
\rho_{\Lambda\lambda}(\vec b_\perp,\vec k_\perp,x)\equiv\frac{1}{2}\left[\rho^{[\gamma^+]}(\vec b_\perp,\vec k_\perp,x,\Lambda\vec e_z)+\lambda\,\rho^{[\gamma^+\gamma_5]}(\vec b_\perp,\vec k_\perp,x,\Lambda\vec e_z)\right].
\end{equation}
We decompose it as follows
\begin{multline}\label{WD}
\rho_{\Lambda\lambda}(\vec b_\perp,\vec k_\perp,x)\\
=\frac{1}{2}\left[\rho_{UU}(\vec b_\perp,\vec k_\perp,x)+\Lambda\,\rho_{LU}(\vec b_\perp,\vec k_\perp,x)+\lambda\,\rho_{UL}(\vec b_\perp,\vec k_\perp,x)+\Lambda\lambda\,\rho_{LL}(\vec b_\perp,\vec k_\perp,x)\right],
\end{multline}
where 
\begin{equation}
\rho_{UU}(\vec b_\perp,\vec k_\perp,x)=\frac{1}{2}\left[\rho^{[\gamma^+]}(\vec b_\perp,\vec k_\perp,x,+\vec e_z)+\rho^{[\gamma^+]}(\vec b_\perp,\vec k_\perp,x,-\vec e_z)\right]
\end{equation}
is the Wigner distribution of unpolarized quarks in an unpolarized nucleon;
\begin{equation}
\rho_{LU}(\vec b_\perp,\vec k_\perp,x)=\frac{1}{2}\left[\rho^{[\gamma^+]}(\vec b_\perp,\vec k_\perp,x,+\vec e_z)-\rho^{[\gamma^+]}(\vec b_\perp,\vec k_\perp,x,-\vec e_z)\right]
\end{equation}
represents the distortion due to the longitudinal polarization of the nucleon;
\begin{equation}
\rho_{UL}(\vec b_\perp,\vec k_\perp,x)=\frac{1}{2}\left[\rho^{[\gamma^+\gamma_5]}(\vec b_\perp,\vec k_\perp,x,+\vec e_z)+\rho^{[\gamma^+\gamma_5]}(\vec b_\perp,\vec k_\perp,x,-\vec e_z)\right]
\end{equation}
represents the distortion due to the longitudinal polarization of the quarks, and
\begin{equation}
\rho_{LL}(\vec b_\perp,\vec k_\perp,x)=\frac{1}{2}\left[\rho^{[\gamma^+\gamma_5]}(\vec b_\perp,\vec k_\perp,x,+\vec e_z)-\rho^{[\gamma^+\gamma_5]}(\vec b_\perp,\vec k_\perp,x,-\vec e_z)\right]
\end{equation}
represents the distortion due to the correlation between quark and nucleon longitudinal polarizations. These four contributions can be written as
\begin{subequations}\label{rhos}
\begin{align}
\rho_{UU}(\vec b_\perp,\vec k_\perp,x)&=
\mathcal{F}_{1,1}(x,0,\vec k_\perp^2,\vec k_\perp\cdot\vec b_\perp,\vec b_\perp^2),\label{UU}\\
\rho_{LU}(\vec b_\perp,\vec k_\perp,x)&=
-\frac{1}{M^2}\,\epsilon_\perp^{ij}k^i_\perp\frac{\partial}{\partial b^j_\perp}\,
\mathcal{F}_{1,4}(x,0,\vec k_\perp^2,\vec k_\perp\cdot\vec b_\perp,\vec b_\perp^2),\label{LU}\\
\rho_{UL}(\vec b_\perp,\vec k_\perp,x)&=
\frac{1}{M^2}\,\epsilon_\perp^{ij}k^i_\perp\frac{\partial}{\partial b^j_\perp}\,
\mathcal{G}_{1,1}(x,0,\vec k_\perp^2,\vec k_\perp\cdot\vec b_\perp,\vec b_\perp^2),\label{UL}\\
\rho_{LL}(\vec b_\perp,\vec k_\perp,x)&=\mathcal{G}_{1,4}(x,0,\vec k_\perp^2,\vec k_\perp\cdot\vec b_\perp,\vec b_\perp^2),\label{LL}
\end{align}
\end{subequations}
where the distributions $\mathcal X=\mathcal{F}_{1,1},\mathcal{F}_{1,4},\mathcal{G}_{1,1},\mathcal{G}_{1,4}$ are the Fourier transforms of the corresponding GTMDs $X=F_{1,1},F_{1,4},G_{1,1},G_{1,4}$ introduced in Ref.~\cite{Meissner:2009ww}
\begin{equation}
\mathcal X(x,\xi,\vec k_\perp^2,\vec k_\perp\cdot\vec b_\perp,\vec b_\perp^2)=\int\frac{\ud^2\Delta_\perp}{(2\pi)^2}\,e^{-i\vec\Delta_\perp\cdot\vec b_\perp}\,X(x,\xi,\vec k_\perp^2,\vec k_\perp\cdot\vec\Delta_\perp,\vec\Delta_\perp^2).
\end{equation}
In Eq.~\eqref{rhos} the two-dimensional antisymmetric tensor $\epsilon^{ij}_\perp$ has been used with $\epsilon^{12}=-\epsilon^{21}=1$, $M$ is the nucleon mass and roman indices are to be summed over. Integrating out $\vec b_\perp$ or $\vec k_\perp$ kills  the contributions $\rho_{LU}$ and $\rho_{UL}$, showing that there exists no TMD or GPD corresponding to $F_{1,4}$ and $G_{1,1}$. These GTMDs carry therefore completely new information about the nucleon structure. On the other hand, the contributions $\rho_{UU}$ and $\rho_{LL}$ survive both integrations. It follows that the GTMD $F_{1,1}$ can be seen as the \emph{mother} distribution of the TMD $f_1$ and the GPD $H$
\begin{subequations}\label{reduction}
\begin{align}
f_1(x,\vec k_\perp^2)&=\int\ud^2b_\perp\,\mathcal{F}_{1,1}(x,0,\vec k_\perp^2,\vec k_\perp\cdot\vec b_\perp,\vec b_\perp^2)=F_{1,1}(x,0,\vec k_\perp^2,0,0),\\
H(x,0,\vec\Delta_\perp^2)&=\int\ud^2k_\perp\,F_{1,1}(x,0,\vec k_\perp^2,\vec k_\perp\cdot\vec \Delta_\perp,\vec\Delta_\perp^2),
\end{align}
and the GTMD $G_{1,4}$ as the \emph{mother} distribution of the TMD $g_{1L}$ and the GPD $\tilde H$
\begin{align}
g_{1L}(x,\vec k_\perp^2)&=\int\ud^2b_\perp\,\mathcal{G}_{1,4}(x,0,\vec k_\perp^2,\vec k_\perp\cdot\vec b_\perp,\vec b_\perp^2)=G_{1,4}(x,0,\vec k_\perp^2,0,0),\\
\tilde H(x,0,\vec\Delta_\perp^2)&=\int\ud^2k_\perp\,G_{1,4}(x,0,\vec k_\perp^2,\vec k_\perp\cdot\vec \Delta_\perp,\vec\Delta_\perp^2).
\end{align} 
\end{subequations}
Integrating out all the variables, one naturally gets
\begin{subequations}
\begin{align}
\int\ud x\,\ud^2k_\perp\,\ud^2b_\perp\,\rho^q_{UU}(\vec b_\perp,\vec k_\perp,x)&=N^q,\\
\int\ud x\,\ud^2k_\perp\,\ud^2b_\perp\,\rho^q_{LU}(\vec b_\perp,\vec k_\perp,x)&=0,\label{zeroLU}\\
\int\ud x\,\ud^2k_\perp\,\ud^2b_\perp\,\rho^q_{UL}(\vec b_\perp,\vec k_\perp,x)&=0,\label{zeroUL}\\
\int\ud x\,\ud^2k_\perp\,\ud^2b_\perp\,\rho^q_{LL}(\vec b_\perp,\vec k_\perp,x)&=\Delta q,
\label{intLL}
\end{align}
\end{subequations}
where the index $q$ indicates the contribution of the quark of flavor $q$,
$N^q$ is the valence-quark number  ($N^u=2$ and $N^d=1$ in the proton) 
and $\Delta q$ is the axial charge.
Note that Eq.~\eqref{zeroLU} tells us that the valence-quark number does not 
depend on the nucleon polarization and Eq.~\eqref{zeroUL} means that in an unpolarized nucleon there is no net quark polarization.

\subsection{Quark Orbital Angular Momentum}
\label{section-2d}

Quantifying quark orbital angular momentum (OAM) inside the nucleon is essential in order to solve the so-called ``spin crisis'', see \emph{e.g.} \cite{Filippone:2001ux,Kuhn:2008sy}. Almost 15 years ago, Ji derived a sum rule that allows one to extract the total quark contribution to the nucleon spin from a combination of GPDs \cite{Ji:1996ek}
\begin{equation}\label{JiJz}
J^q_{z}=\frac{1}{2}\int\ud x\,x\left[H^q(x,0,0)+E^q(x,0,0)\right].
\end{equation}
By subtracting half of the axial charge $\Delta q=\int\ud x\,\tilde H^q(x,0,0)$ which is interpreted as the spin contribution of quarks with flavor $q$ to the nucleon spin, one gets the quark OAM contribution
\begin{equation}\label{JiLz}
L^q_{z}=\frac{1}{2}\int\ud x\left\{x\left[H^q(x,0,0)+E^q(x,0,0)\right]-\tilde H^q(x,0,0)\right\}.
\end{equation}
From a density point of view, this result is surprising in the sense that the extraction of the quark OAM along the $z$-axis involves the GPD $E$ which appears only in a transversely polarized nucleon. Note however that $E$ describes the amplitude where the nucleon spin flips while the quark light-cone helicities remain unaffected, implying therefore a change by one unit of OAM between the initial and final nucleon states.

More recently it has been suggested, based on some quark models, that the TMD $h_{1T}^\perp$ may also be related to the quark OAM \cite{Avakian:2008dz,Avakian:2010br,Avakian:2009jt,She:2009jq}
\begin{equation}\label{pretzelosity}
\mathcal L^q_{z}=-\int\ud x\,\ud^2k_\perp\,\frac{\vec k_\perp^2}{2M^2}\,h_{1T}^{\perp q}(x,\vec k_\perp^2).
\end{equation}
Note that one expects in general $\mathcal L^q_{z}\neq L^q_{z}$ in a gauge theory, see \emph{e.g.} \cite{Burkardt:2008ua}. Once again, from a density point of view, this expression is surprising in the sense that it involves the TMD $h_{1T}^\perp$ which describes the distribution of transversely polarized quarks in a transversely polarized nucleon. Note however that $h_{1T}^\perp$ corresponds to the amplitude where the nucleon and active quark longitudinal polarizations flip in opposite directions, involving therefore a change by two units of OAM between the initial and final nucleon states.

Clearly, Wigner distributions provide much more information than GPDs and TMDs as they contain also the full correlations between quark transverse position and three-momentum. Furthermore, once the Wigner distributions are known, it is rather straightforward to compute  physical observables. One has just to take the phase-space average as if the Wigner distributions were classical distributions
\begin{equation}
\langle\widehat A\rangle^{[\Gamma]}(\vec S)=\int\ud x\,\ud^2k_\perp\,\ud^2b_\perp\,A(\vec b_\perp,\vec k_\perp,x)\,\rho^{[\Gamma]}(\vec b_\perp,\vec k_\perp,x,\vec S).
\end{equation}
In particular, we can write the average quark OAM in a nucleon polarized in the $z$-direction as
\begin{eqnarray}
\ell^q_{z}\equiv\langle\widehat L^q_z\rangle^{[\gamma^+]}(\vec e_z)&=&
\int\ud x\,\ud^2k_\perp\,\ud^2b_\perp\left(\vec b_\perp\times\vec k_\perp\right)_z\rho^{[\gamma^+]q}(\vec b_\perp,\vec k_\perp,x,\vec e_z)\nonumber\\
&=&\int\ud x\,\ud^2k_\perp\,\ud^2b_\perp\left(\vec b_\perp\times\vec k_\perp\right)_z
\,[\rho^{q}_{UU}(\vec b_\perp,\vec k_\perp,x)
+ \rho^{q}_{LU}(\vec b_\perp,\vec k_\perp,x)].
\end{eqnarray}
 From Eq.~\eqref{UU}, it is clear that
\begin{equation}\label{zeroUU}
\int\ud x\,\ud^2k_\perp\,\ud^2b_\perp\left(\vec b_\perp\times\vec k_\perp\right)_z\rho^q_{UU}(\vec b_\perp,\vec k_\perp,x)=0,
\end{equation}
which means that in an unpolarized nucleon there is no net quark OAM\footnote{An unpolarized nucleon has no spin, which means that the total quark and gluon angular momentum contributions have to sum up to zero. By rotational invariance, one expects all the four contributions (spin and OAM of quarks and gluons) to vanish identically. The angular momentum sum rule for an unpolarized nucleon is therefore trivially satisfied.}. Using now Eq.~\eqref{LU} and integrating by parts, we find that the quark OAM $\ell^q_{z}$ reads
\begin{equation}\label{LzGTMD}
\ell^q_{z}=-\int\ud x\,\ud^2k_\perp\,\frac{\vec k_\perp^2}{M^2}\,F_{1,4}^q(x,0,\vec k_\perp^2,0,0).
\end{equation}
An interesting issue which deserves further investigation is the relation between $L^q_{z}$ in Eq.~(\ref{JiLz}) and $\ell^q_{z}$ in Eq.~\eqref{LzGTMD}. As discussed in the following sections, in models without gauge-field degrees of freedom one finds that the two definitions give the same results for the total quark contribution to the OAM, but not for the separate quark-flavor contributions. However, this remains to be confirmed in more complex systems, when the contribution of the Wilson line is explicitly taken into account.

Wigner distributions allow us also to study the correlation between quark spin and OAM, which we define as
\begin{equation}\label{corrSpinOAM}
\begin{split}
C^q_{z}&\equiv\int\ud x\,\ud^2k_\perp\,\ud^2b_\perp\left(\vec b_\perp\times\vec k_\perp\right)_z\rho^q_{UL}(\vec b_\perp,\vec k_\perp,x)\\
&=\int\ud x\,\ud^2k_\perp\,\frac{\vec k_\perp^2}{M^2}\,G^q_{1,1}(x,0,\vec k_\perp^2,0,0),
\end{split}
\end{equation}
where we have used Eq.~\eqref{UL}. For $C^q_{z}>0$ the quark spin and OAM tend to be aligned, while for $C^q_{z}<0$ they tend to be antialigned. Finally, note that from Eq.~\eqref{LL} one has
\begin{equation}
\int\ud x\,\ud^2k_\perp\,\ud^2b_\perp\left(\vec b_\perp\times\vec k_\perp\right)_z\rho_{LL}(\vec b_\perp,\vec k_\perp,x)=0,
\label{OAMLL}
\end{equation}
reflecting like Eqs.~\eqref{zeroLU}, \eqref{zeroUL} and \eqref{zeroUU} the isotropy of space.

\section{Results and Discussions}
\label{section-3}

Since it is not known how to extract Wigner distributions or GTMDs from experiments, one has to rely on phenomenological models. We studied the Wigner distributions in the light-cone constituent quark model (LCCQM) \cite{Boffi:2002yy,Pasquini:2005dk,Pasquini:2008ax} and the light-cone version of the chiral quark-soliton model ($\chi$QSM) restricted to the three-quark sector \cite{Petrov:2002jr,Diakonov:2005ib,Lorce:2006nq,Lorce:2007as,Lorce:2007fa}, using the general formalism developed in Ref.~\cite{Lorce:2011dv} for the overlap representation of the quark-quark correlator in terms of light-cone wave functions. We neglect  the contribution from gauge degrees of freedom, and in particular from the Wilson line  in the Wigner operator~\eqref{wigner-operator}. As the resulting distributions are very similar in both models, we will present only those from the LCCQM. However, when discussing more quantitative aspects, we will also report the numerical values from the $\chi$QSM. Furthermore, we will discuss only the first $x$ moment of the Wigner distributions
\begin{equation}
\rho(\vec b_\perp,\vec k_\perp)\equiv\int\ud x\,\rho(\vec b_\perp,\vec k_\perp,x),
\end{equation}
\emph{i.e.} purely transverse four-dimensional phase-space distributions (two transverse position and two transverse momentum coordinates), referred to as \emph{transverse} Wigner distributions.

\subsection{Unpolarized Quarks in an Unpolarized Nucleon}
\label{section-3a}

\begin{figure}[th!]
	\centering
		\includegraphics[width=.49\textwidth]{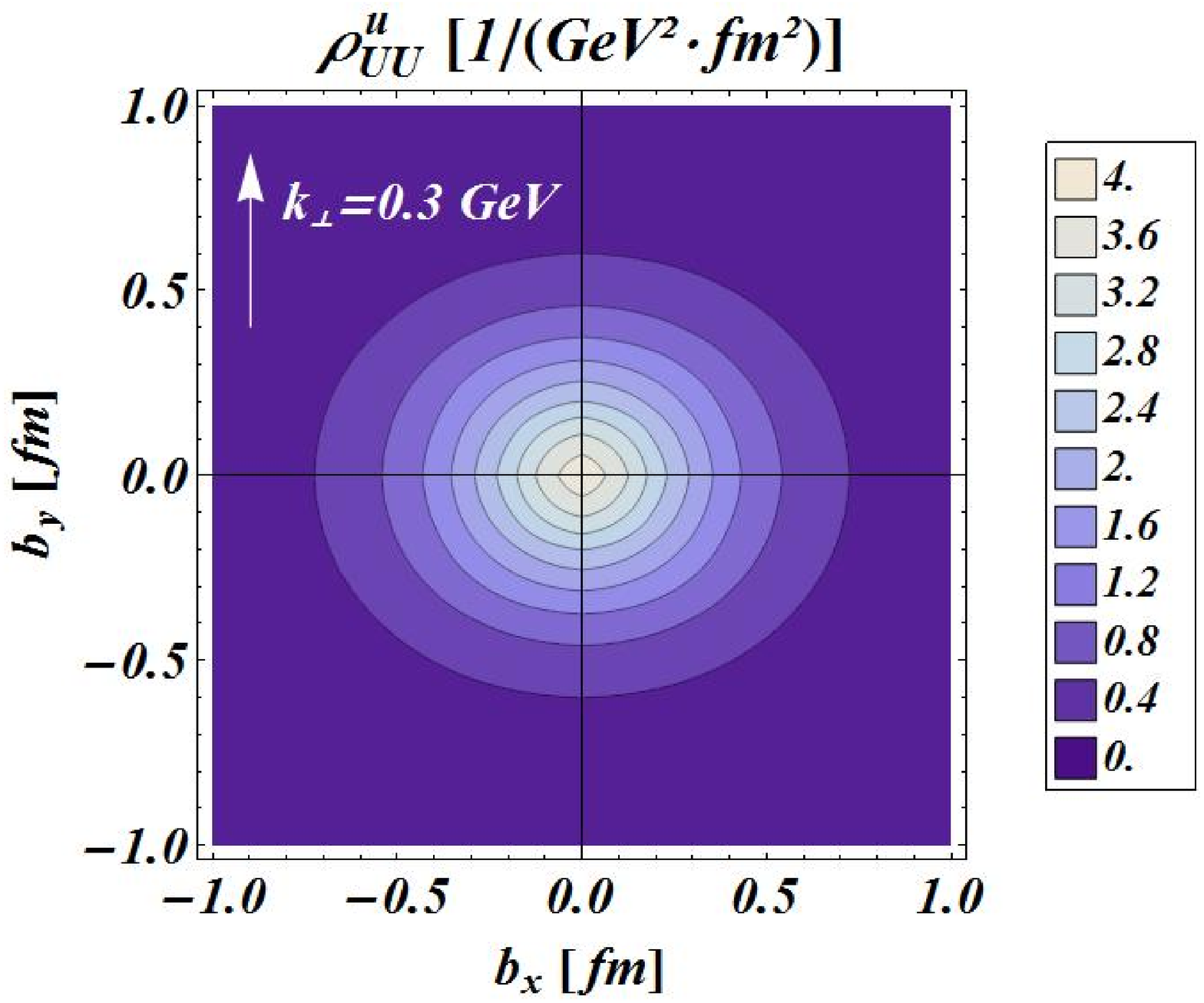}
		\includegraphics[width=.49\textwidth]{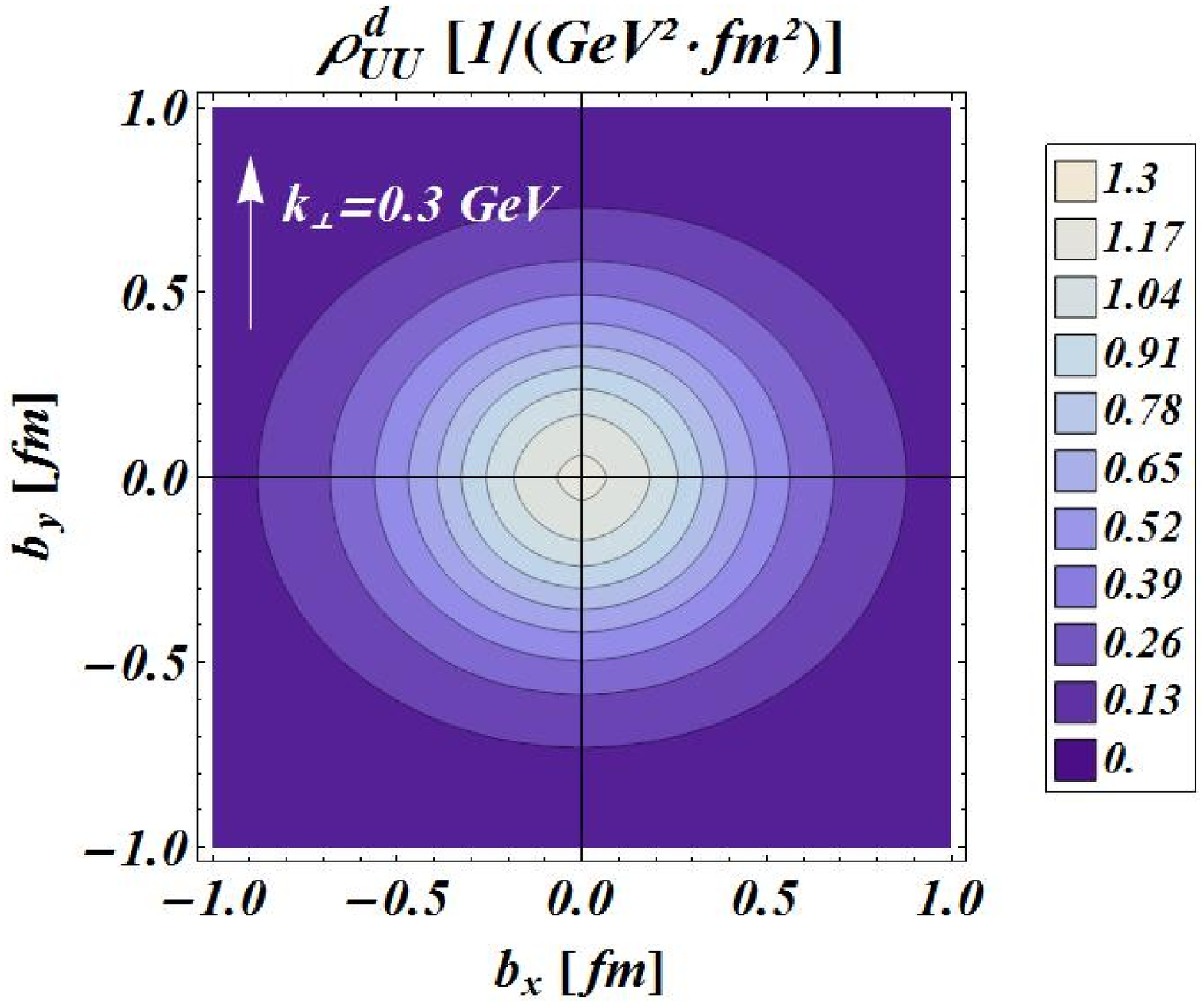}
		\includegraphics[width=.49\textwidth]{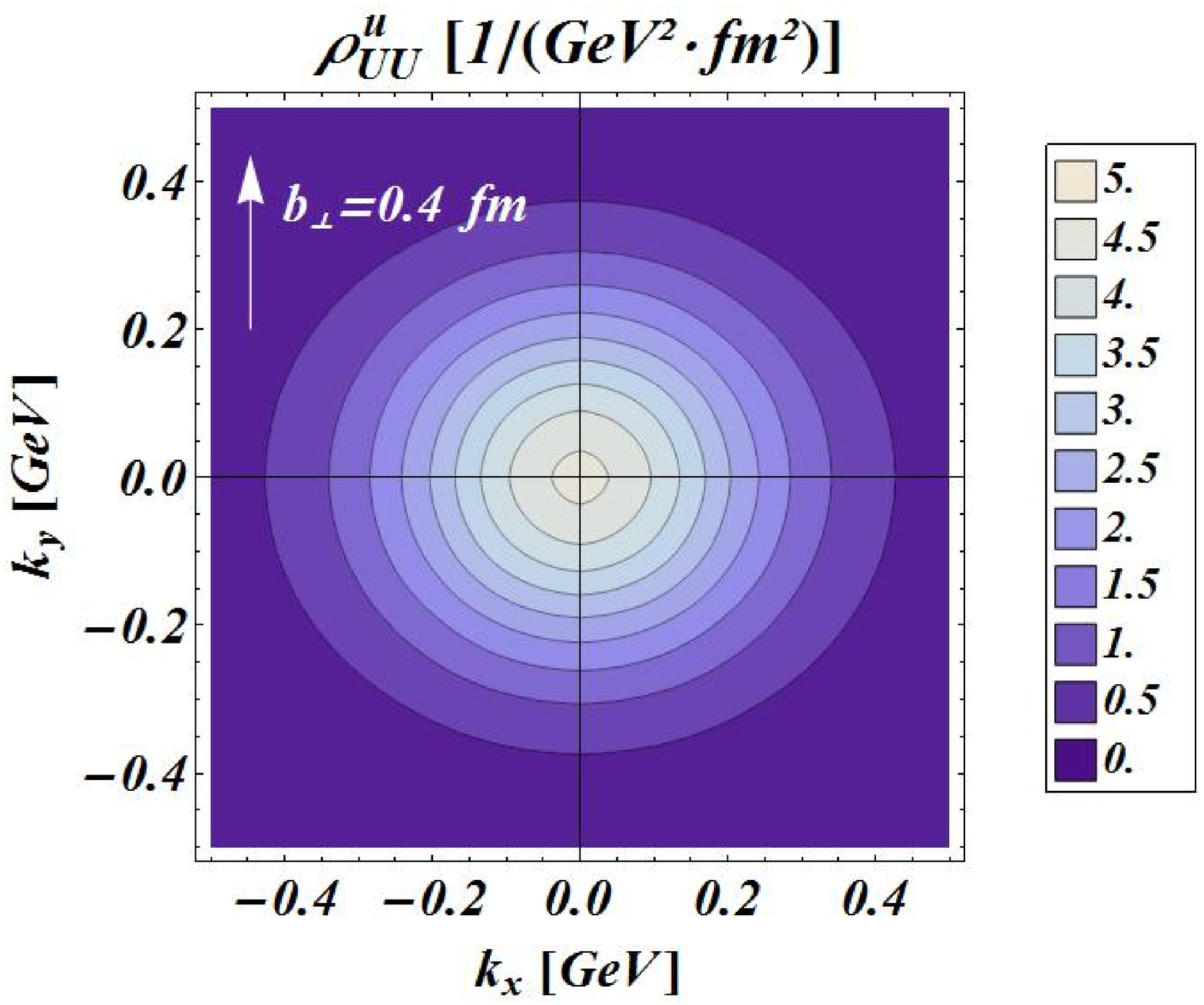}
		\includegraphics[width=.49\textwidth]{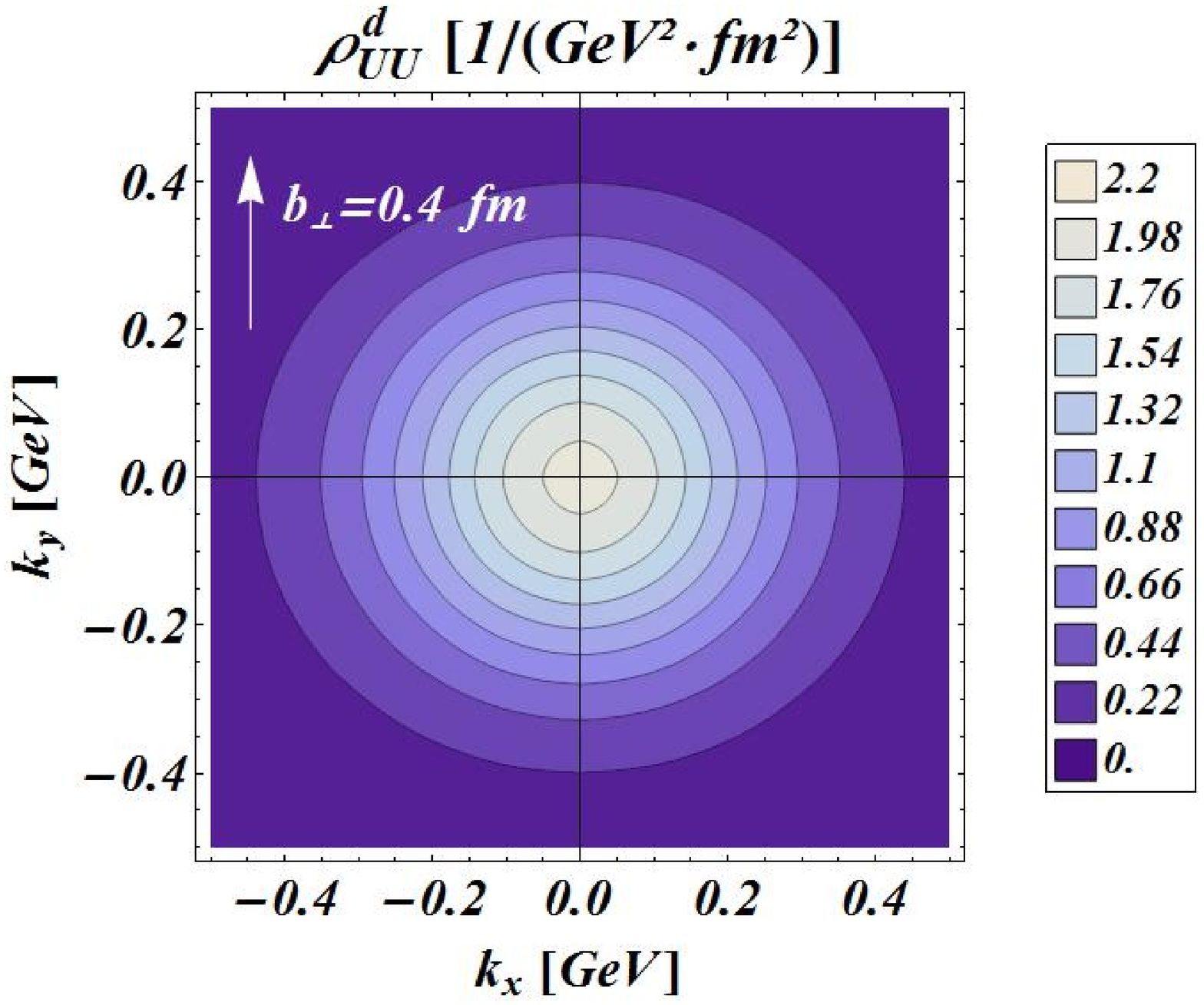}
\caption{\footnotesize{The transverse Wigner distributions of unpolarized  quarks in an unpolarized proton. Upper panels:  distributions in impact-parameter space with fixed transverse momentum $\vec k_\perp=k_\perp\,\hat e_y$ and $k_\perp=0.3$ GeV. Lower panels: distributions in transverse-momentum space with fixed impact parameter $\vec b_\perp=b_\perp\,\hat e_y$ and $b_\perp=0.4$ fm. 
The left (right) panels show the results for $u$ ($d$) quarks.}}
		\label{fig1}
\end{figure}

We start the discussions with $\rho_{UU}(\vec b_\perp,\vec k_\perp)$, the transverse Wigner distribution of unpolarized quarks in an unpolarized proton. In Fig.~\ref{fig1} we present the distributions in impact-parameter space  with fixed transverse momentum $\vec k_\perp=k_\perp\,\hat e_y$ and $k_\perp=0.3$ GeV (upper panels), and compare them with the distribution in transverse-momentum space with fixed impact parameter $\vec b_\perp=b_\perp\,\hat e_y$ and $b_\perp=0.4$ fm (lower panels).
The left (right) panels refer to the $u$ ($d$) quarks. We observe a distortion in all these distributions which indicates that the configuration $\vec b_\perp\perp\vec k_\perp$ is favored with respect to the configuration $\vec b_\perp\parallel\vec k_\perp$. This can be understood with naive semi-classical arguments. The radial momentum $(\vec k_\perp\cdot\hat b_\perp)\,\hat b_\perp$ ($\hat b_\perp\equiv\vec b_\perp/b_\perp$) of a quark is expected  to decrease rapidly in the periphery because of confinement. The polar momentum $\vec k_\perp-(\vec k_\perp\cdot\hat b_\perp)\,\hat b_\perp$ receives a contribution from the orbital motion of the quark which can still be significant in the periphery (in an orbital motion, one does not need to reduce the momentum to avoid a quark escape). This naive picture also suggests that this phenomenon should be stronger as we go to peripheral regions ($b_\perp\gg$) and to high quark momenta ($k_\perp\gg$). Such a behavior is indeed observed in our model calculations and can be quantified in terms of the average quadrupole distortions $Q_b^{ij}(\vec k_\perp)$ and $Q_k^{ij}(\vec b_\perp)$ defined as
\begin{subequations}\label{Quad}
\begin{align}
Q_b^{ij}(\vec k_\perp)&=Q_b(k_\perp)\left(2\hat k^i\hat k^j-\delta^{ij}\right)=
\frac{\int{\rm d}^2b_\perp\left(2 b^{i}_\perp b^{j}_\perp-\delta^{ij} b^2_\perp\right)\rho_{UU}(\vec b_\perp,\vec k_\perp)}
{\int{\rm d}^2b_\perp\,b^2_\perp\,\rho_{UU}(\vec b_\perp,\vec k_\perp)},\\
Q_k^{ij}(\vec b_\perp)&=Q_k(b_\perp)\left(2\hat b^i\hat b^j-\delta^{ij}\right)
=\frac{\int{\rm d}^2k_\perp\left(2k^i_\perp k^j_\perp-\delta^{ij}k^2_\perp\right)
\rho_{UU}(\vec b_\perp,\vec k_\perp)}
{\int{\rm d}^2k_\perp\,k^2_\perp\,\rho_{UU}(\vec b_\perp,\vec k_\perp)},
\end{align}
\end{subequations}
where $i,j=x,y$. The distortions calculated in the LCCQM are tabulated in Table~\ref{quadrupoletable}.
\begin{table}[th!]
\begin{center}
\caption{\footnotesize{The average quadrupole distortions of the transverse Wigner distribution of unpolarized quarks in an unpolarized proton from the LCCQM. See Eq.~\eqref{Quad} for the definition of $Q_b(k_\perp)$ and $Q_k(b_\perp)$.}}\label{quadrupoletable}
\begin{tabular}{@{\quad}c@{\quad}|@{\quad}c@{\quad}c@{\quad}||@{\quad}c@{\quad}|@{\quad}c@{\quad}c@{\quad}}\whline
\multirow{2}{*}{$k_\perp$ [GeV]}&\multicolumn{2}{@{\quad}c@{\quad}||@{\quad}}{$Q_b(k_\perp)$}&\multirow{2}{*}{$b_\perp$ [fm]}&\multicolumn{2}{@{\quad}c@{\quad}}{$Q_k(b_\perp)$}\\
&$u$&$d$&&$u$&$d$\\
\hline
$0$&$0$&$0$&$0$&$0$&$0$\\
$0.1$&$-0.04$&$-0.03$&$0.2$&$-0.15$&$-0.10$\\
$0.2$&$-0.14$&$-0.13$&$0.4$&$-0.24$&$-0.19$\\
$0.3$&$-0.30$&$-0.27$&$0.6$&$-0.29$&$-0.27$\\
\whline
\end{tabular}
\end{center}
\end{table} 
We note that the quadrupole distortions of $u$ and $d$ quarks are very similar. For increasing values of $k_\perp$ and $b_\perp$ the distortions get more and more pronounced, and at the same time the spread of the distributions shrinks towards the center. From Fig.~\ref{fig1} we also note that the spread of the distributions is smaller for $u$ quarks than for $d$ quarks, especially in the transverse-coordinate space. This reflects the fact that $u$ quarks are more concentrated at the center of the proton, while the $d$-quark distribution has a tail which extends further at the periphery of the proton.

From Eq.~\eqref{UU} we see that $\rho_{UU}(\vec b_\perp,\vec k_\perp)=\rho_{UU}(b_\perp,k_\perp,\vec k_\perp\cdot\vec b_\perp)$. This explains the left-right symmetry in Fig.~\ref{fig1} and implies that the quark is as likely to rotate clockwise as to rotate anticlockwise.  In Fig.~\ref{fig1} we also observe a top-bottom symmetry. Such a symmetry is not a general property of the Wigner distribution $\rho_{UU}$, but follows from the fact that in our calculations there are no explicit gluons. Indeed, time-reversal invariance implies that the real part of the GTMDs is $T$-even ($e$) while the imaginary part is $T$-odd ($o$). Hermiticity tells us that the four GTMDs $X=F_{1,1},F_{1,4},G_{1,1},G_{1,4}$ satisfy the relations
\begin{align*}
X^e(x,\xi,\vec k_\perp^2,\vec k_\perp\cdot\vec\Delta_\perp,\vec\Delta_\perp^2)&=X^e(x,-\xi,\vec k_\perp^2,-\vec k_\perp\cdot\vec\Delta_\perp,\vec\Delta_\perp^2),\\
X^o(x,\xi,\vec k_\perp^2,\vec k_\perp\cdot\vec\Delta_\perp,\vec\Delta_\perp^2)&=-X^o(x,-\xi,\vec k_\perp^2,-\vec k_\perp\cdot\vec\Delta_\perp,\vec\Delta_\perp^2).
\end{align*}
This means that for $\xi=0$, $X^e$ is an even function of $\vec k_\perp\cdot\vec\Delta_\perp$ while $X^o$ is an odd function of $\vec k_\perp\cdot\vec\Delta_\perp$. It follows that the Fourier transforms $\mathcal X$ of these GTMDs with respect to $\vec\Delta_\perp$ are real-valued functions. The contribution $\mathcal X^e$ is an even function of $\vec k_\perp\cdot\vec b_\perp$ while $\mathcal X^o$ is an odd function of $\vec k_\perp\cdot\vec b_\perp$. Since we have no explicit gluons and therefore no final-state interactions, our GTMDs are real. It follows that $\mathcal X=\mathcal X^e$ explaining the top-bottom symmetry of Fig.~\ref{fig1}. The dominant effect of final-state interactions would be to shift up or down the distributions.

\begin{figure}[th!]
	\centering
		\includegraphics[width=.49\textwidth]{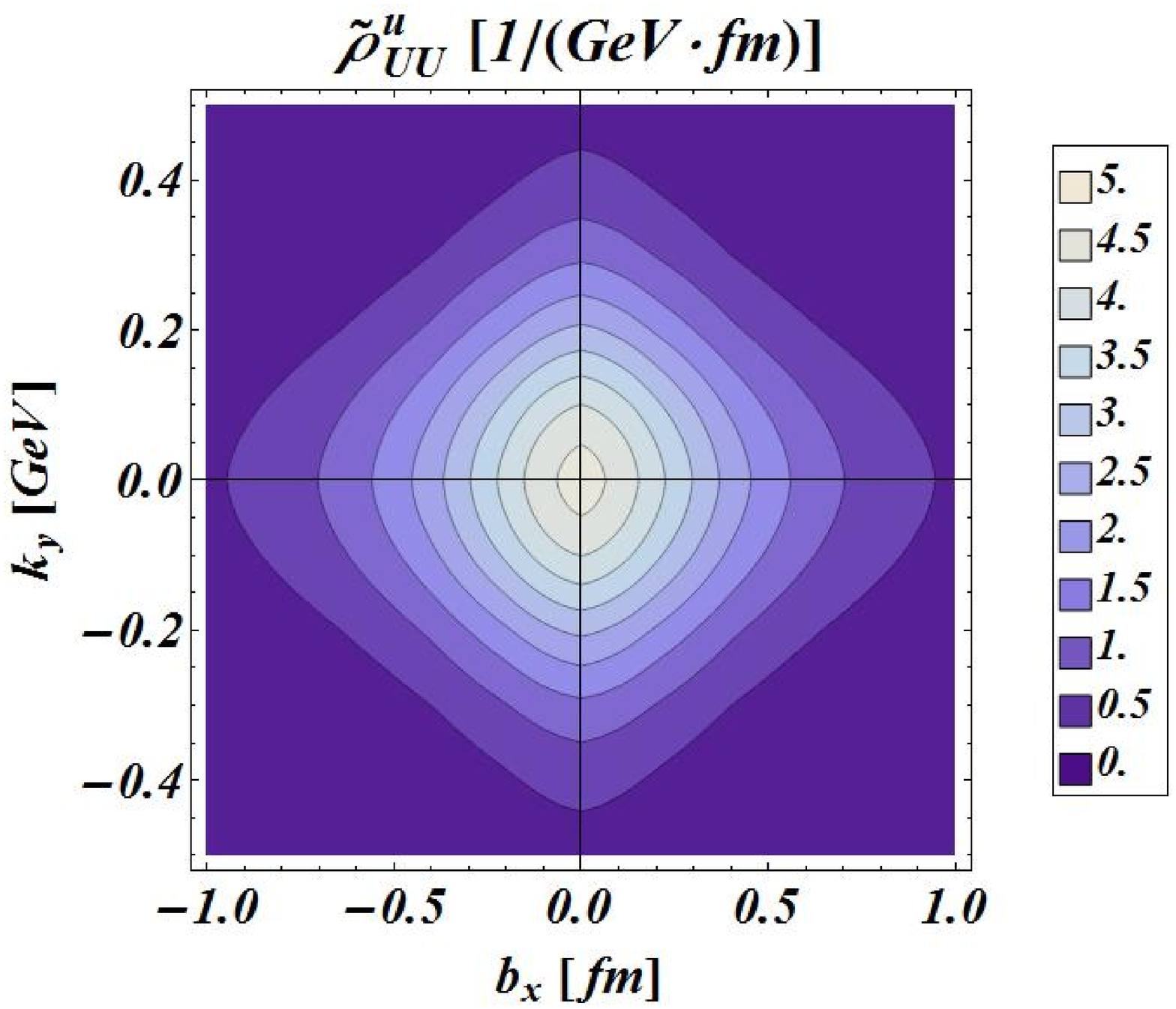}
		\includegraphics[width=.49\textwidth]{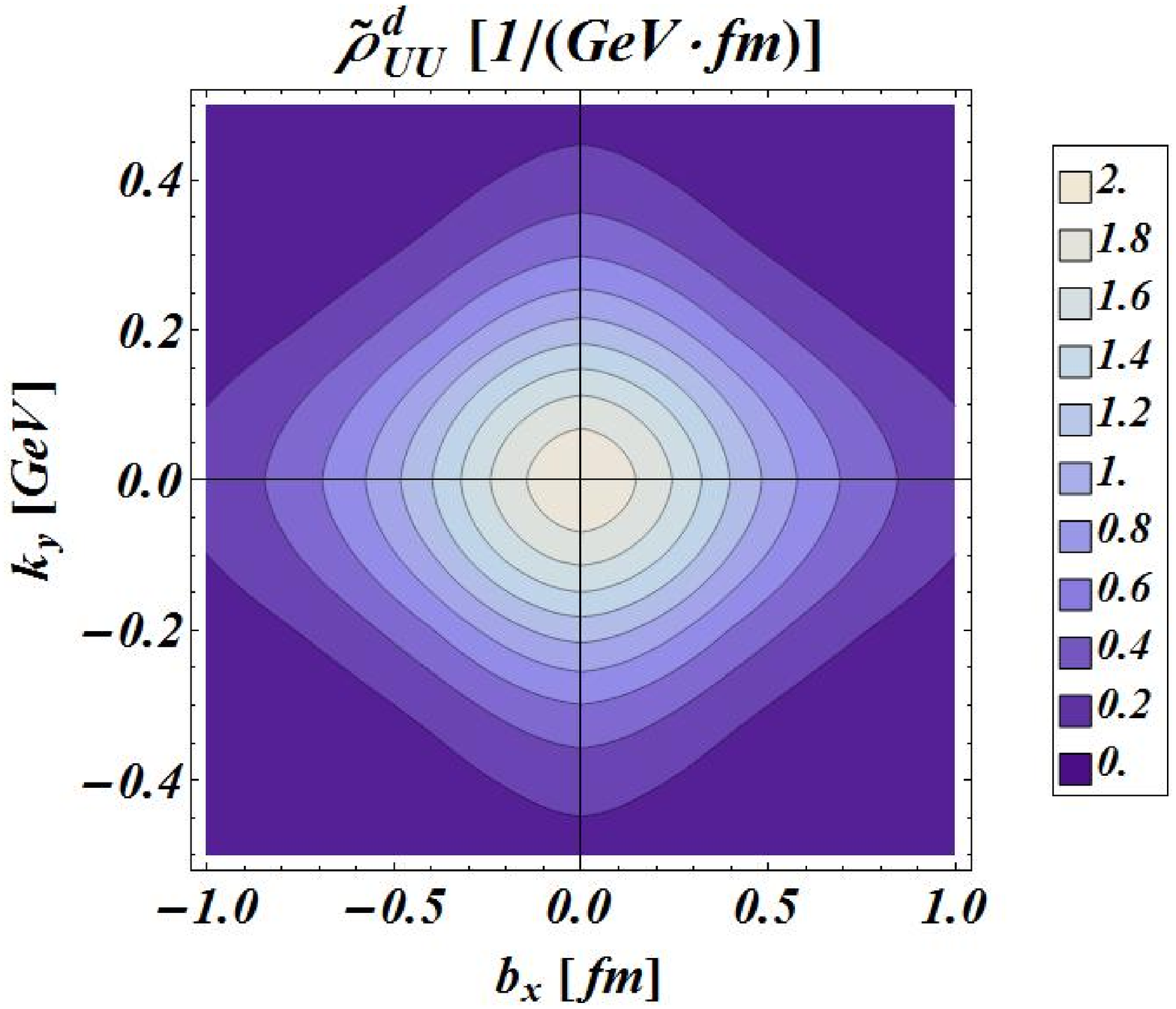}
		\caption{\footnotesize{The mixed transverse densities $\tilde\rho(b_x,k_y)$} of unpolarized $u$ quarks (left panel) and unpolarized $d$ quarks (right panel) in an unpolarized proton.}
		\label{fig3}
\end{figure}
As mentioned earlier, Wigner distributions have only a quasi-probabilistic interpretation due to Heisenberg uncertainty relations. A genuine probabilistic interpretation can be recovered only when integrating out certain variables.  If we integrate out $\vec b_\perp$ or $\vec k_\perp$, we reduce to the unpolarized TMD and GPD, respectively. In these cases, the distortion we observed in the Wigner distribution $\rho_{UU}$ is completely absent and we are left with axially symmetric distributions, see Eq.~\eqref{reduction}. By integrating over one momentum and one coordinate variables which are not conjugated, we obtain the probability densities $\tilde\rho$ of Eqs.~\eqref{bxky} and \eqref{bykx}. In Fig.~\ref{fig3} we show the probability density $\tilde\rho_{UU}(b_x,k_y)$ which gives the correlation between $b_x$ and $k_y$. We observe that $\tilde \rho_{UU}(b_x,k_y)$ is maximum at the center, $b_x=k_y=0$, and decreases in the outer regions of the phase space, with the equidensity lines in each quadrant of Fig.~\ref{fig3} having approximately a linear dependence in $b_x$ and $k_y$. Furthermore, we clearly see that the width of the densities in $k_y$ is similar for $u$ and $d$ quarks, while it is more extended in $b_x$ for $d$ quarks than for $u$ quarks.

\subsection{Unpolarized Quarks in a Longitudinally Polarized Nucleon}
\label{section-3b}

\begin{figure}[th!]
	\centering
		\includegraphics[width=.49\textwidth]{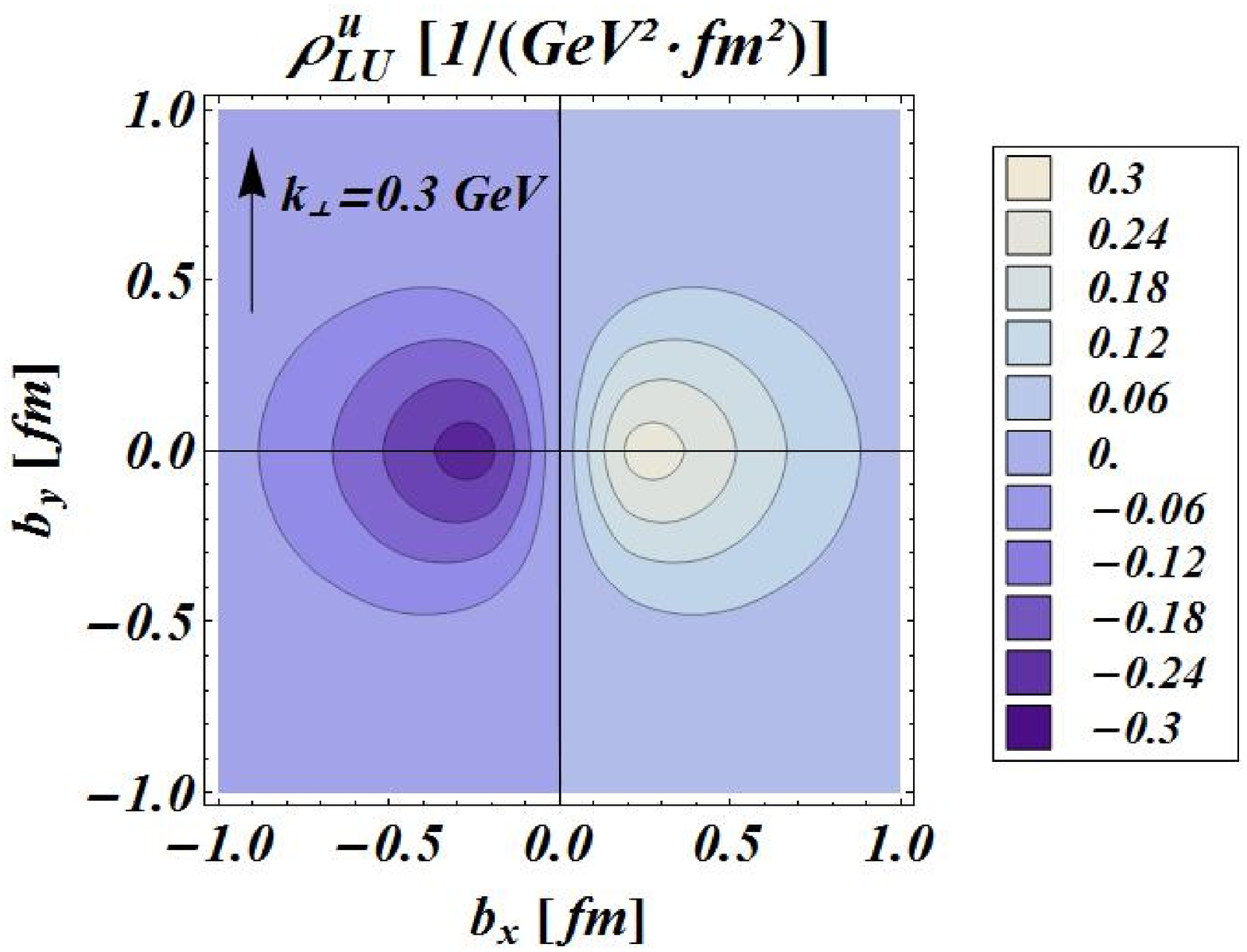}
		\includegraphics[width=.49\textwidth]{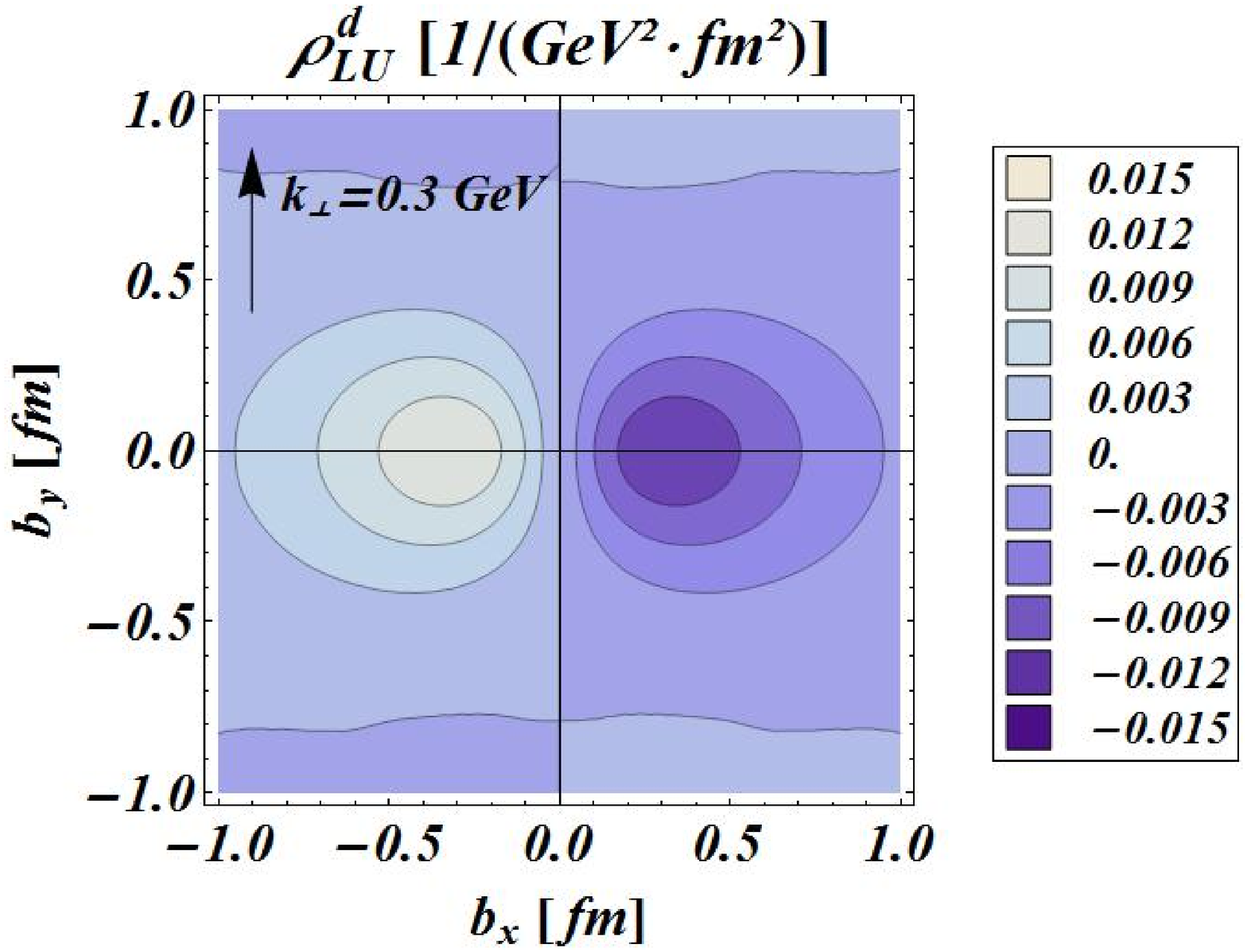}
		\includegraphics[width=.49\textwidth]{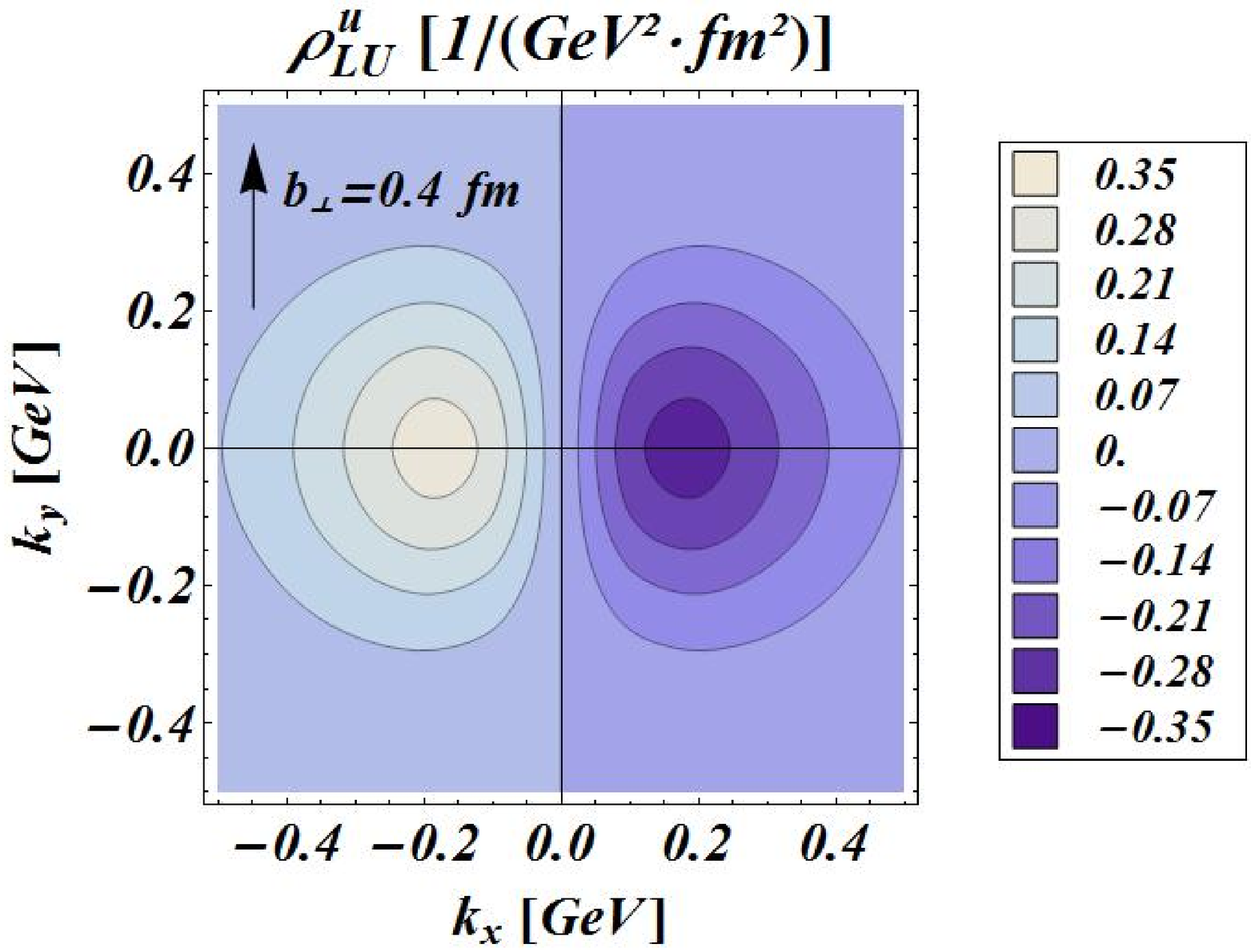}
		\includegraphics[width=.49\textwidth]{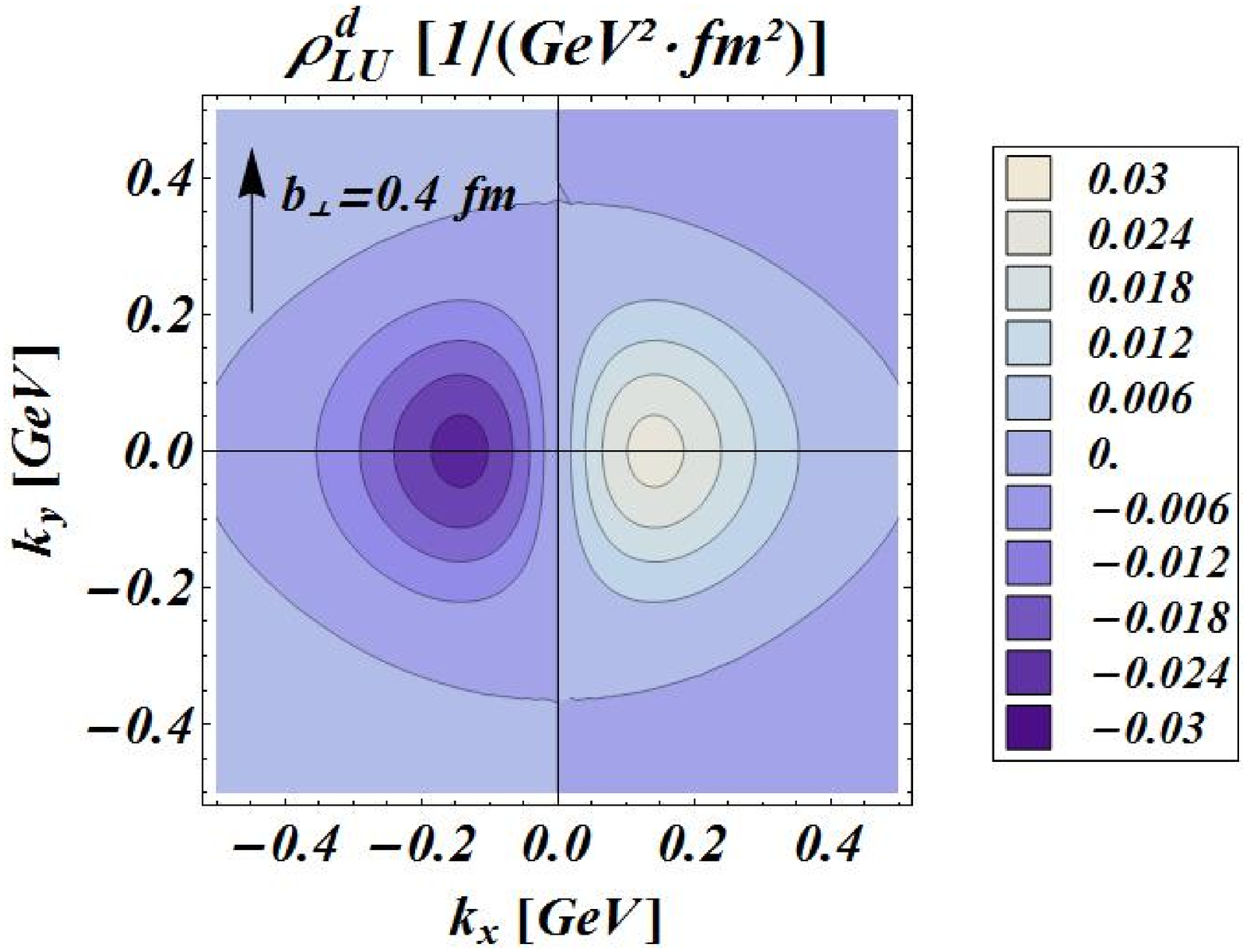}
		\caption{\footnotesize{The distortions
 of the transverse Wigner distributions of unpolarized quarks due to 
the spin of the proton (pointing out of the plane). Upper panels:  distortions in impact-parameter space with fixed transverse momentum $\vec k_\perp=k_\perp\,\hat e_y$ and $k_\perp=0.3$ GeV. Lower panels: distortions in transverse-momentum space with fixed impact parameter $\vec b_\perp=b_\perp\,\hat e_y$ and $b_\perp=0.4$ fm. The left (right) panels show the results for $u$ ($d$) quarks.
}}\label{fig4}
\end{figure}

We now  consider $\rho_{LU}(\vec b_\perp,\vec k_\perp)$, the distortion of the transverse Wigner distribution of unpolarized quarks due to the longitudinal polarization of the proton. In Fig.~\ref{fig4}, the upper panels show the distortions in impact-parameter space for $u$ (left panels) and $d$ (right panels) quarks with fixed transverse momentum $\vec k_\perp=k_\perp\,\hat e_y$ and $k_\perp=0.3$ GeV, while the lower panels give the corresponding distortions in the transverse-momentum space with fixed impact parameter $\vec b_\perp=b_\perp \hat e_y$ and $b_\perp=0.4 $ fm.
\begin{figure}[th!]
	\centering
		\includegraphics[width=.49\textwidth]{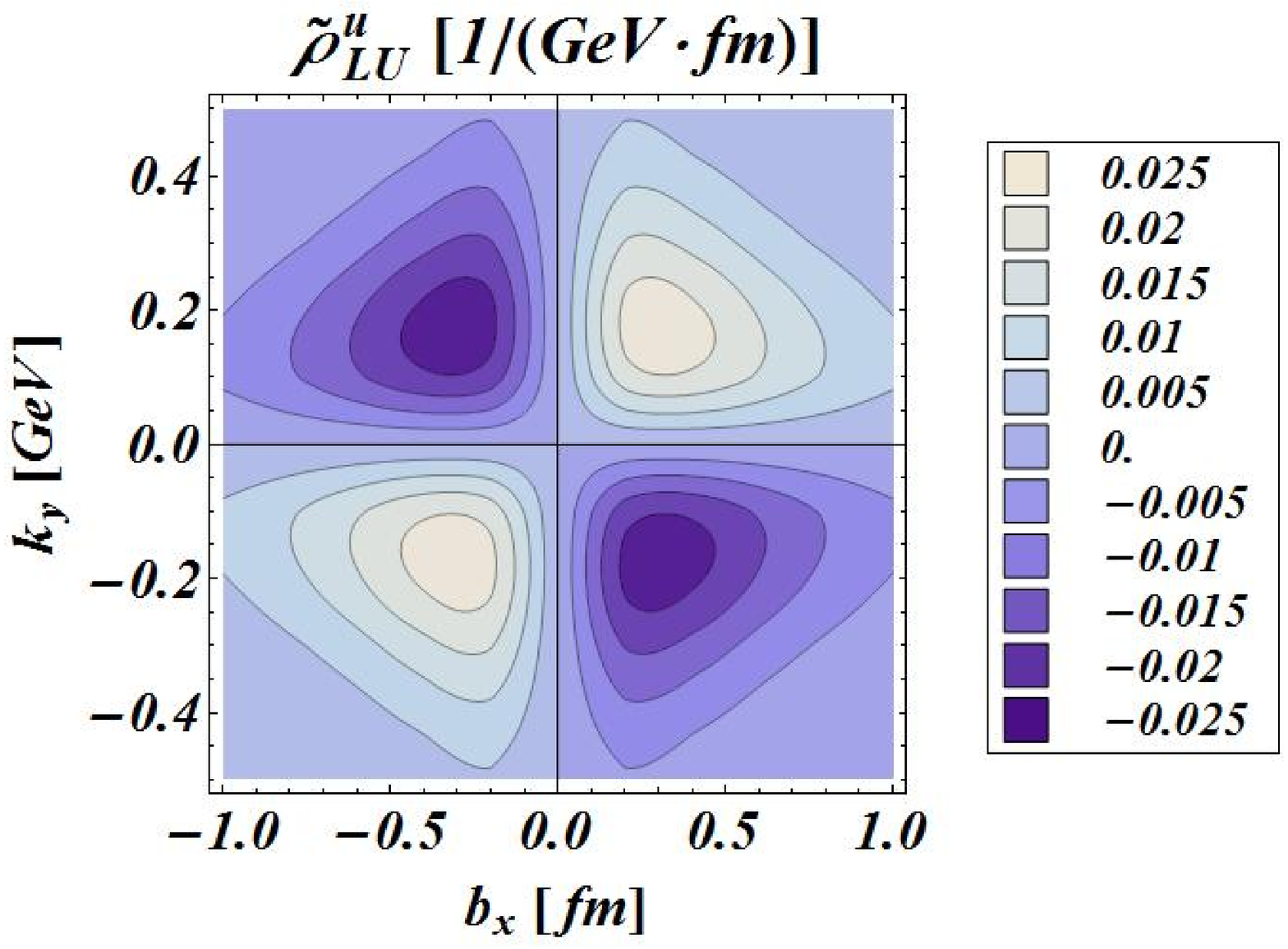}
		\includegraphics[width=.49\textwidth]{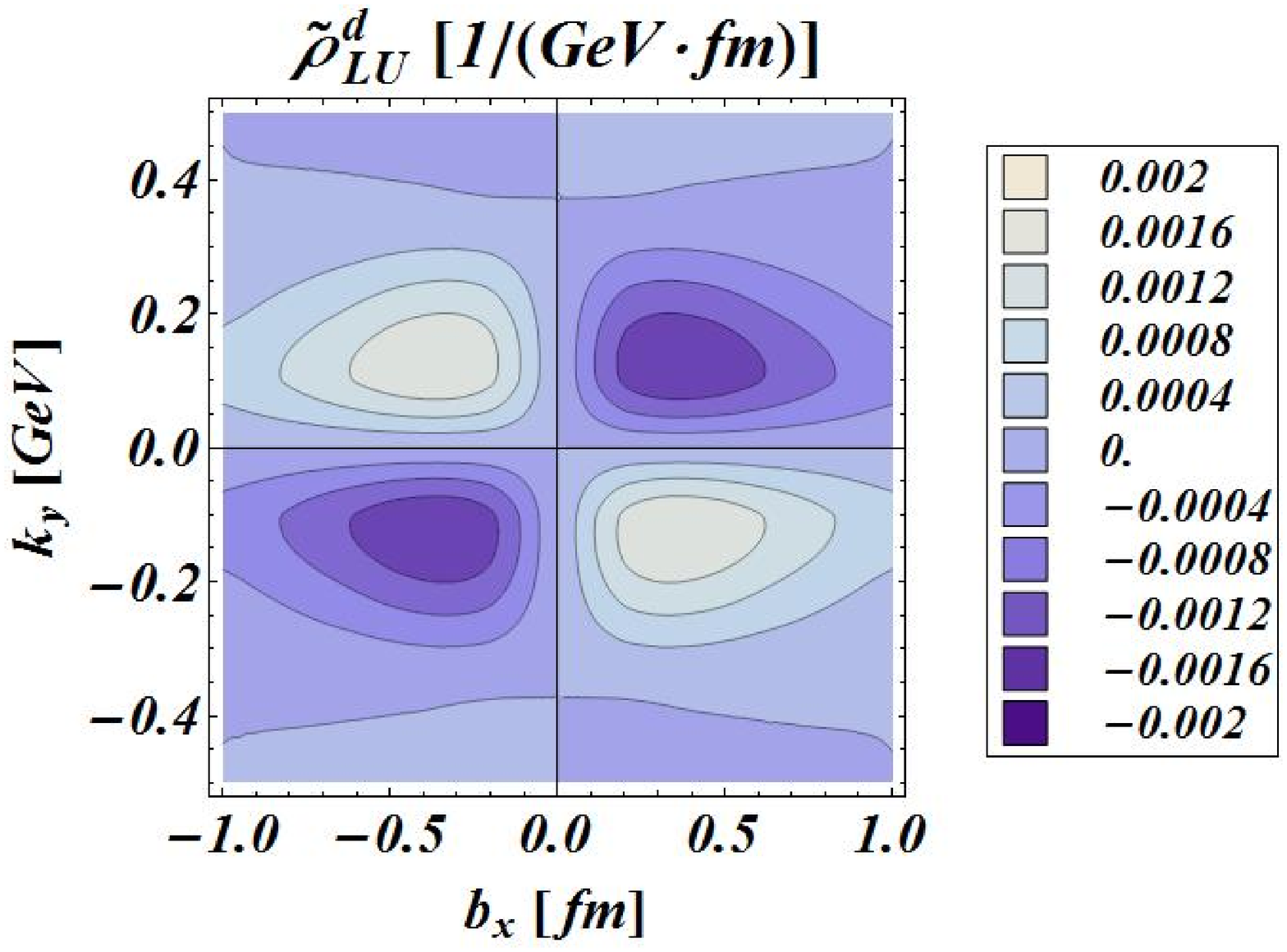}
\caption{\footnotesize{The distortions of the mixed transverse densities $\tilde\rho(b_x,k_y)$ of unpolarized $u$ quarks (left panel) and $d$ quarks (right panel) due to the spin of the proton (pointing out of the plane).}}
		\label{fig5}
\end{figure}
We observe a clear dipole structure in both these distributions, with opposite sign for $u$ and $d$ quarks. The corresponding distortions of the mixed transverse densities $\tilde\rho(b_x,k_y)$ are shown in Fig.~\ref{fig5} for the $u$ (left panel) and $d$ (right panel) quarks. In this case, we observe a quadrupole structure. These multipole structures are due to the explicit factor $\epsilon^{ij}_\perp k^i_\perp\tfrac{\partial}{\partial b_\perp^j}$ in Eq.~\eqref{LU} which breaks the left-right symmetry in Fig.~\ref{fig4} allowing therefore non-vanishing net OAM. We learn from these figures that the OAM of $u$ quarks tends be aligned with the nucleon spin, while the OAM of $d$ quarks tends be antialigned with the nucleon spin. In particular, we notice that the distortion induced by the quark OAM is stronger in the central region of the phase space ($k_\perp\ll$ and $b_\perp\ll$), for both $u$ and $d$ quarks. The distortion in the $\vec b_\perp$ space (see upper panels of Fig.~\ref{fig4}) is more extended for $d$ quarks than for $u$ quarks, whereas the opposite behavior is found for the distortion in the $\vec k_\perp$ space (see lower panels of Fig.~\ref{fig4}). In the case of $d$ quarks, we also observe a sign change of the distributions in the outer regions of phase space ($k_\perp\gg$ and $b_\perp\gg$) which corresponds to a flip of the local net quark OAM.
\begin{table}[th!]
\begin{center}
\caption{\footnotesize{The results for quark orbital angular momentum (see Eqs.~\eqref{JiLz}, \eqref{pretzelosity} and \eqref{LzGTMD}) and anomalous magnetic moment $\kappa^q$ from the LCCQM and the $\chi$QSM for $u$-, $d$- and total ($u+d$) quark contributions.}}\label{OAMtable}
\begin{tabular}{@{\quad}c@{\quad}c@{\quad}|@{\quad}c@{\quad}c@{\quad}c@{\quad}|@{\quad}c@{\quad}c@{\quad}c@{\quad}}\whline
\multicolumn{2}{@{\quad}c@{\quad}|@{\quad}}{Model}&\multicolumn{3}{c@{\quad}|@{\quad}}{LCCQM}&\multicolumn{3}{c@{\quad}}{$\chi$QSM}\\
\multicolumn{2}{@{\quad}c@{\quad}|@{\quad}}{$q$}&$u$&$d$&Total&$u$&$d$&Total\\
\hline
$\ell^q_z$&Eq.~\eqref{LzGTMD}&$0.131$&$-0.005$&$0.126$&$0.073$&$-0.004$&$0.069$\\
$L^q_z$&Eq.~\eqref{JiLz}&$0.071$&$0.055$&$0.126$&$-0.008$&$0.077$&$0.069$\\
$\mathcal L^q_z$&Eq.~\eqref{pretzelosity}&$0.169$&$-0.042$&$0.126$&$0.093$&$-0.023$&$0.069$\\
\hline
$\kappa^q$&&$1.867$&$-1.579$&$0.288$&$1.766$&$-1.551$&$0.215$\\
\whline
\end{tabular}
\end{center}
\end{table}

According to Eq.~\eqref{LzGTMD}, the integral over the phase space of the distribution in Fig.~\ref{fig4} multiplied by $(\vec b_\perp\times \vec k_\perp)_z$ gives the expectation value of the quark OAM. The corresponding results for $u$-, $d$- and total ($u+d$) quark contributions are reported in the first row of Table~\ref{OAMtable}. We give the predictions from both the LCCQM and the $\chi$QSM. They are compared with the corresponding results for the orbital angular momentum $L^q_z$ and $\mathcal L^q_z$ obtained from the definitions in Eqs.~\eqref{JiLz} and \eqref{pretzelosity}, respectively. We note that all the three definitions give the same results for the total quark OAM, but differ for the separate quark-flavor contributions. In agreement with the interpretation of the TMD results given in Ref.~\cite{Lorce:2011dv}, there is more net quark OAM in the LCCQM ($\sum_q L^q_z=0.126$) than in the $\chi$QSM ($\sum_q L^q_z=0.069$). For the individual quark contributions, both the LCCQM and the $\chi$QSM predict that $\ell^q_{z}$ and  $\mathcal L^q_z$  are positive for $u$ quarks and negative for $d$ quarks, with the $u$-quark contribution larger than the $d$-quark contribution in absolute value. For $L_z^q$ the LCCQM predicts the same positive sign for the $u$ and $d$ contributions, with the isovector combination $L^u_z- L^d_z>0$, similarly to a variety of relativistic quark model calculations~\cite{Burkardt:2008ua}. Instead, the $\chi$QSM gives $L_z^u<0$  and $L_z^d>0$, and therefore $L^u_z-L^d_z<0$, in agreement with lattice calculations~\cite{Hagler:2007xi,Hagler:2009ni}. Note however that the quark angular momenta $J^q_z$ of Eq.~\eqref{JiJz} are very similar in both models. We find $J^u_z=0.569$ ($0.566$) and $J^d_z=-0.069$ ($-0.066$) in the LCCQM ($\chi$QSM). This is due to the fact that the difference in $L^q_z$ between the two model predictions is compensated by the different results for the spin contribution, namely $\Delta u=0.995$ ($1.148$) and $\Delta d=-0.249$ ($-0.287$) in the LCCQM ($\chi$QSM).

In the last row of Table~\ref{OAMtable} we also give the results for the quark anomalous magnetic moment $\kappa^q$ which is intimately connected to quark OAM. In particular, it is well known within the light-cone wave function description of hadrons that a state can have anomalous magnetic moment only in the presence of nonzero OAM components~\cite{Brodsky:1980zm,Brodsky:2000ii,Burkardt:2005km,Lu:2006kt}. Furthermore, the anomalous magnetic moment gives a measurement of the correlation between the transverse spin of the nucleon and the orbital motion of quarks, as observed in the IPDs for unpolarized quarks in a transversely polarized nucleon~\cite{Burkardt:2005td,Burkardt:2002hr}. We find the following pattern $0<-\kappa^{d}<\kappa^{u}$ which coincides with the ones for $\mathcal L^q_{z}$ and $\ell^q_{z}$ but not for $L^q_{z}$.

\subsection{Longitudinally Polarized Quarks in an Unpolarized Nucleon}
\label{section-3c}

\begin{figure}[th!]
	\centering
		\includegraphics[width=.49\textwidth]{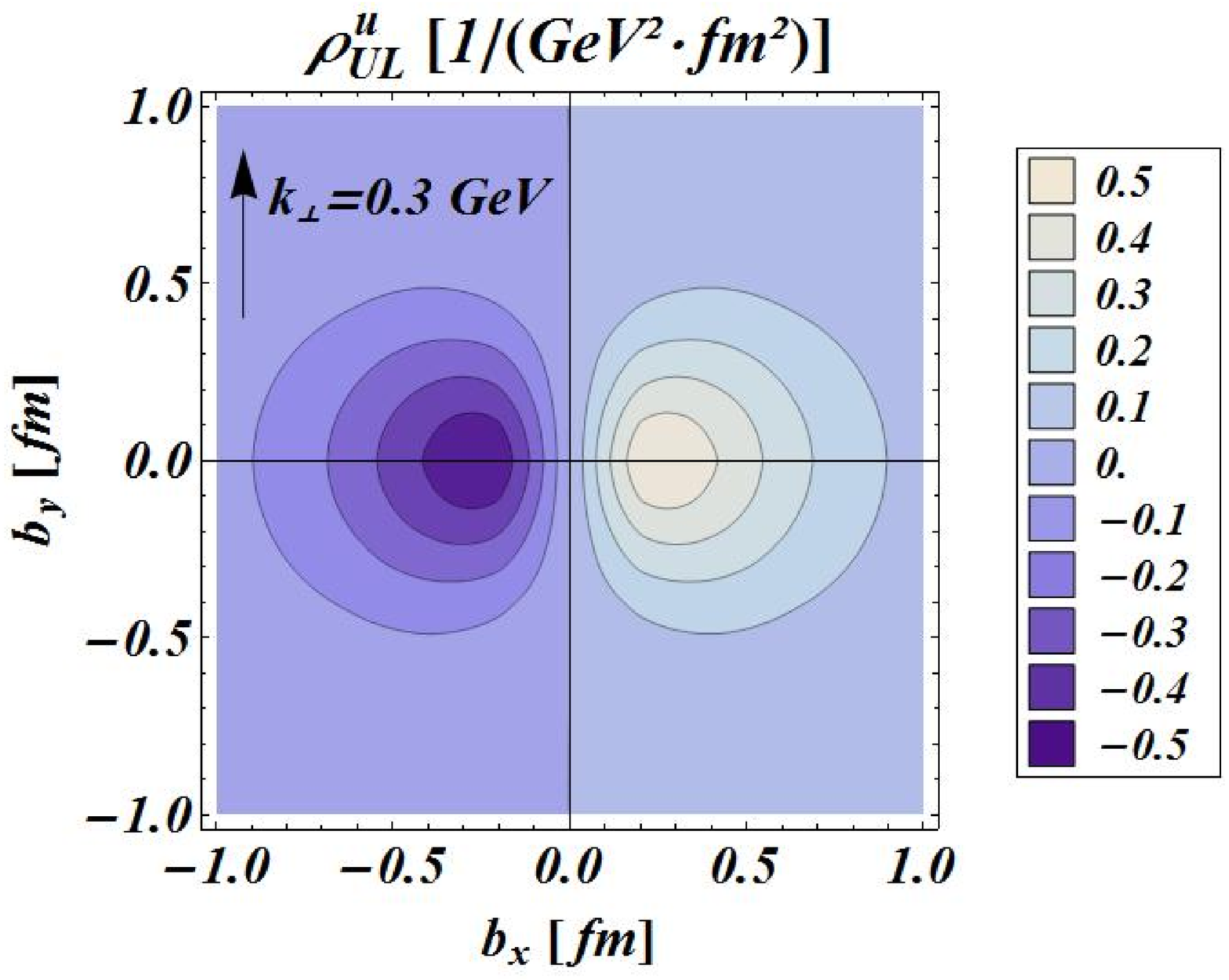}
		\includegraphics[width=.49\textwidth]{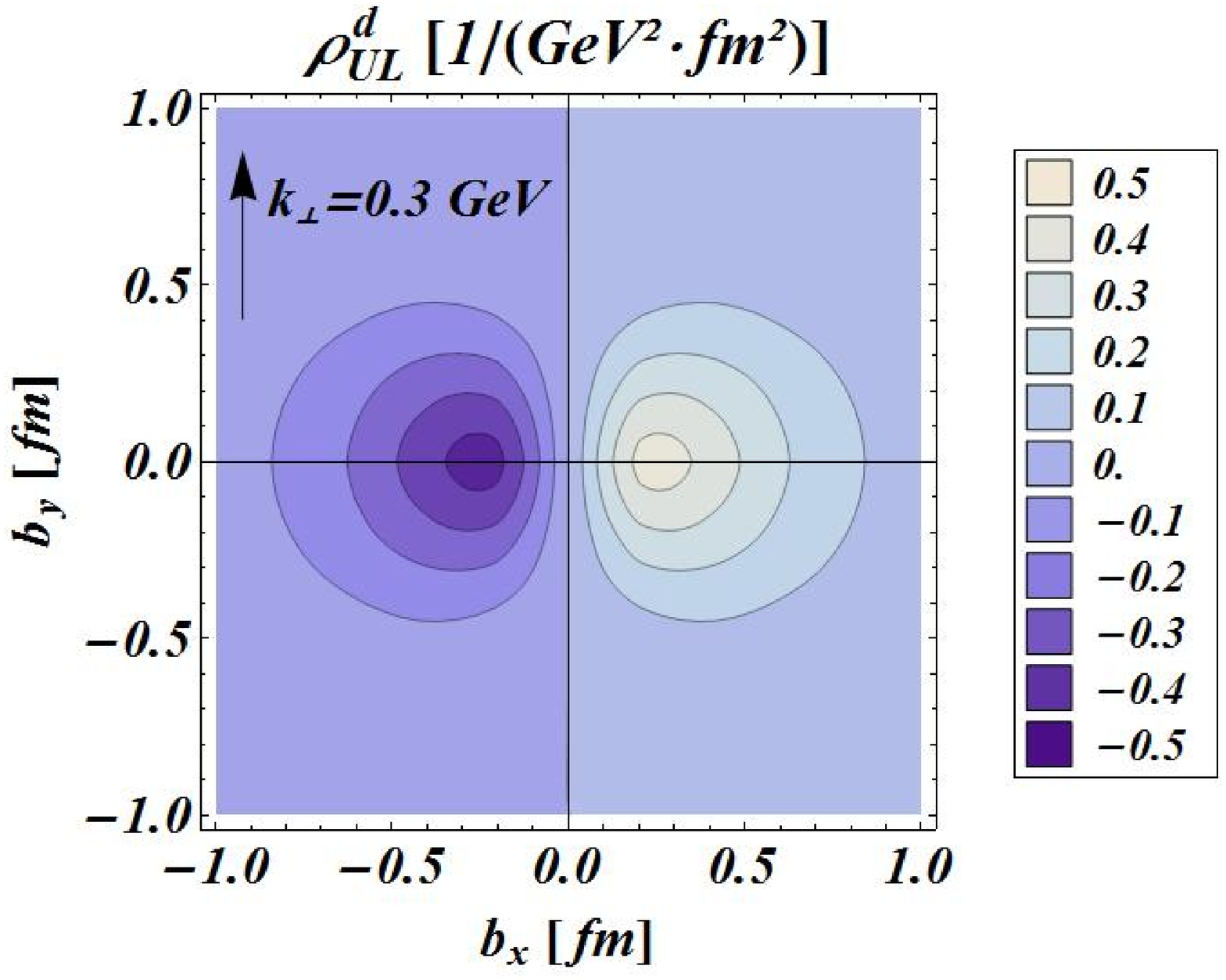}
		\includegraphics[width=.49\textwidth]{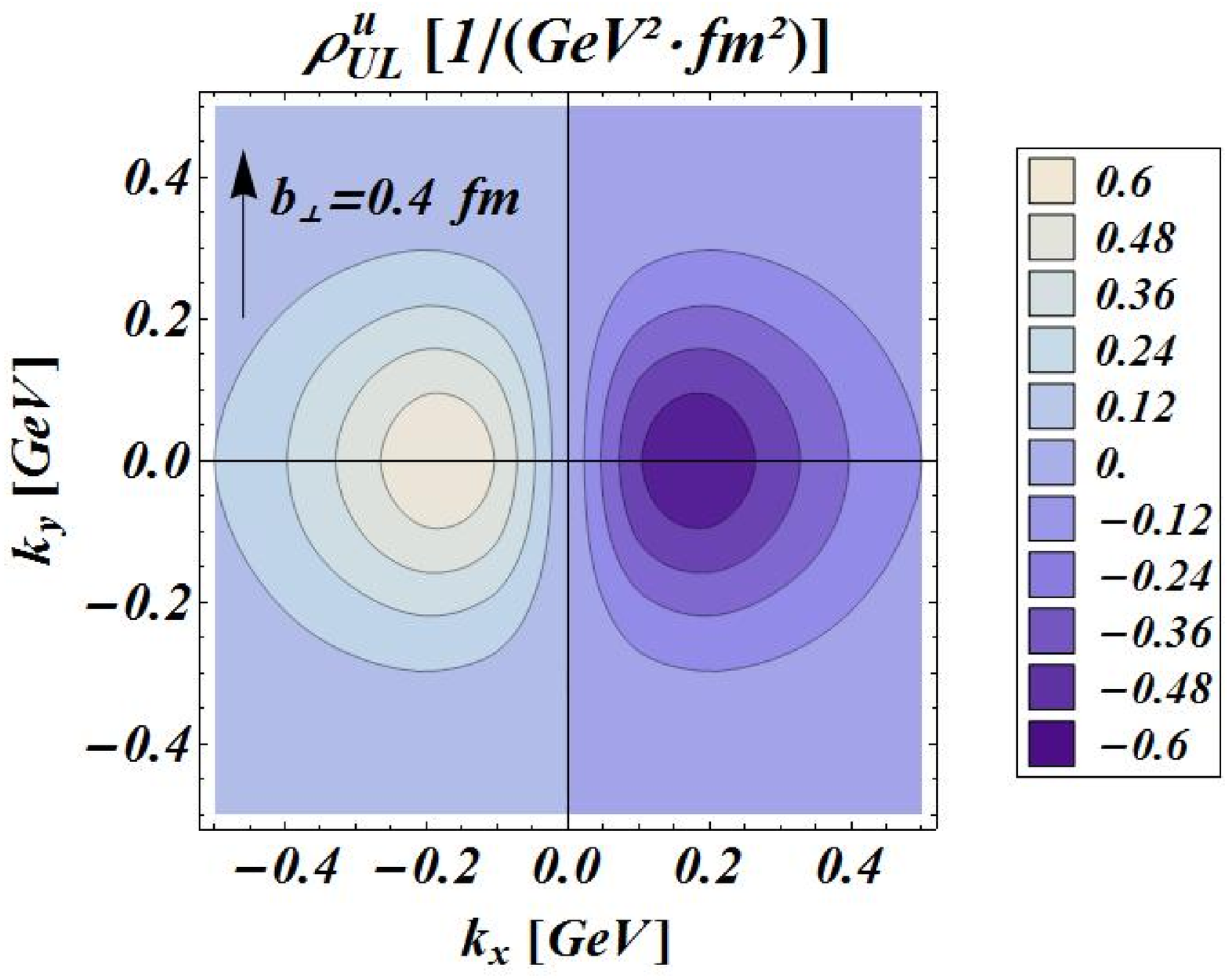}
		\includegraphics[width=.49\textwidth]{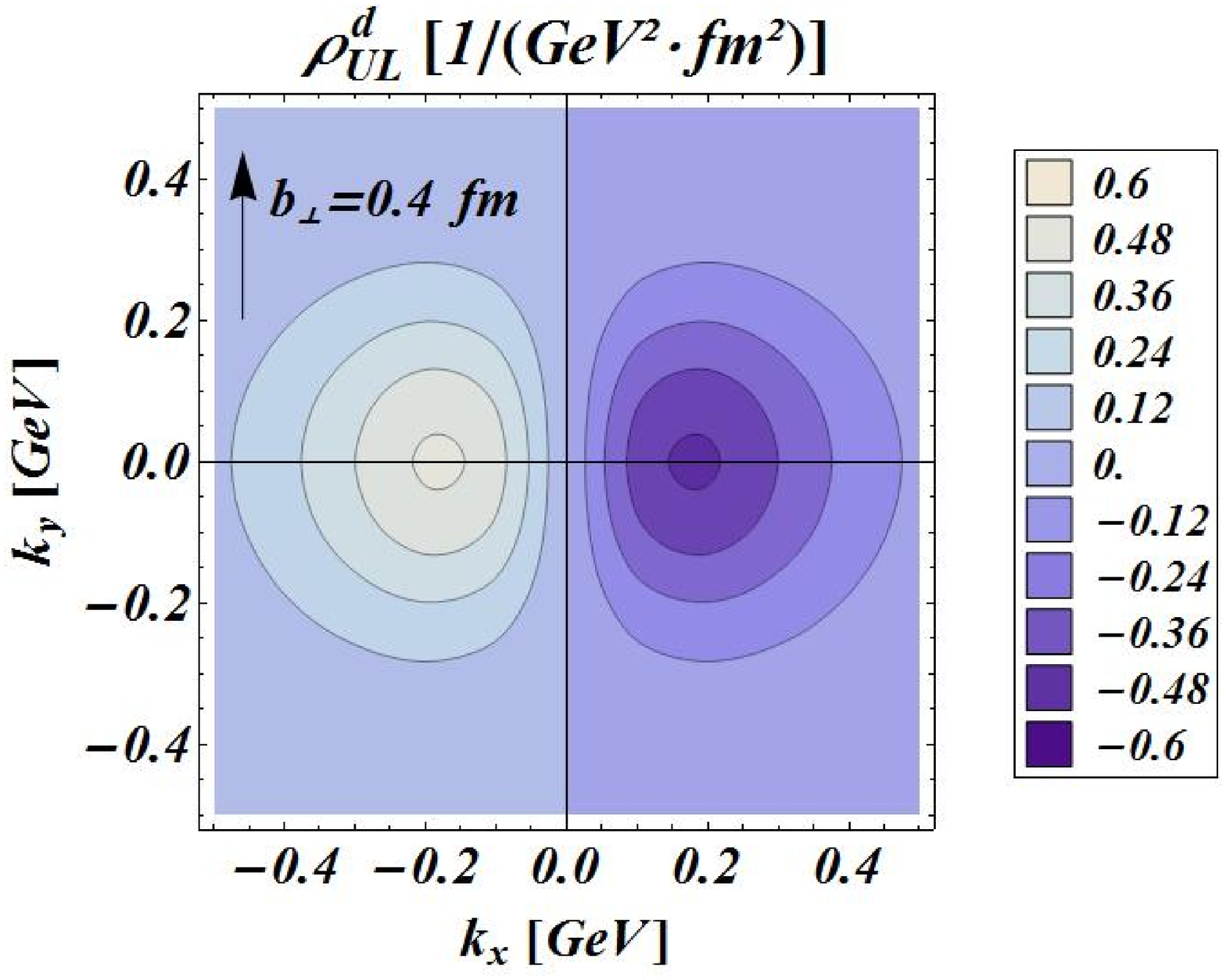}
\caption{\footnotesize{The distortions of the transverse Wigner distributions due to the spin of the quarks (pointing out of the plane) in an unpolarized proton. Upper panels: distortions in impact-parameter space with fixed transverse momentum $\vec k_\perp=k_\perp\,\hat e_y$ and $k_\perp=0.3$ GeV. Lower panels: distortions in transverse-momentum space with fixed impact parameter $\vec b_\perp=b_\perp\,\hat e_y$ and $b_\perp=0.4$ fm. The  left (right) panels show the results for $u$ ($d$) quarks.}}
		\label{fig6}
\end{figure}
\begin{figure}[th!]
	\centering
		\includegraphics[width=.49\textwidth]{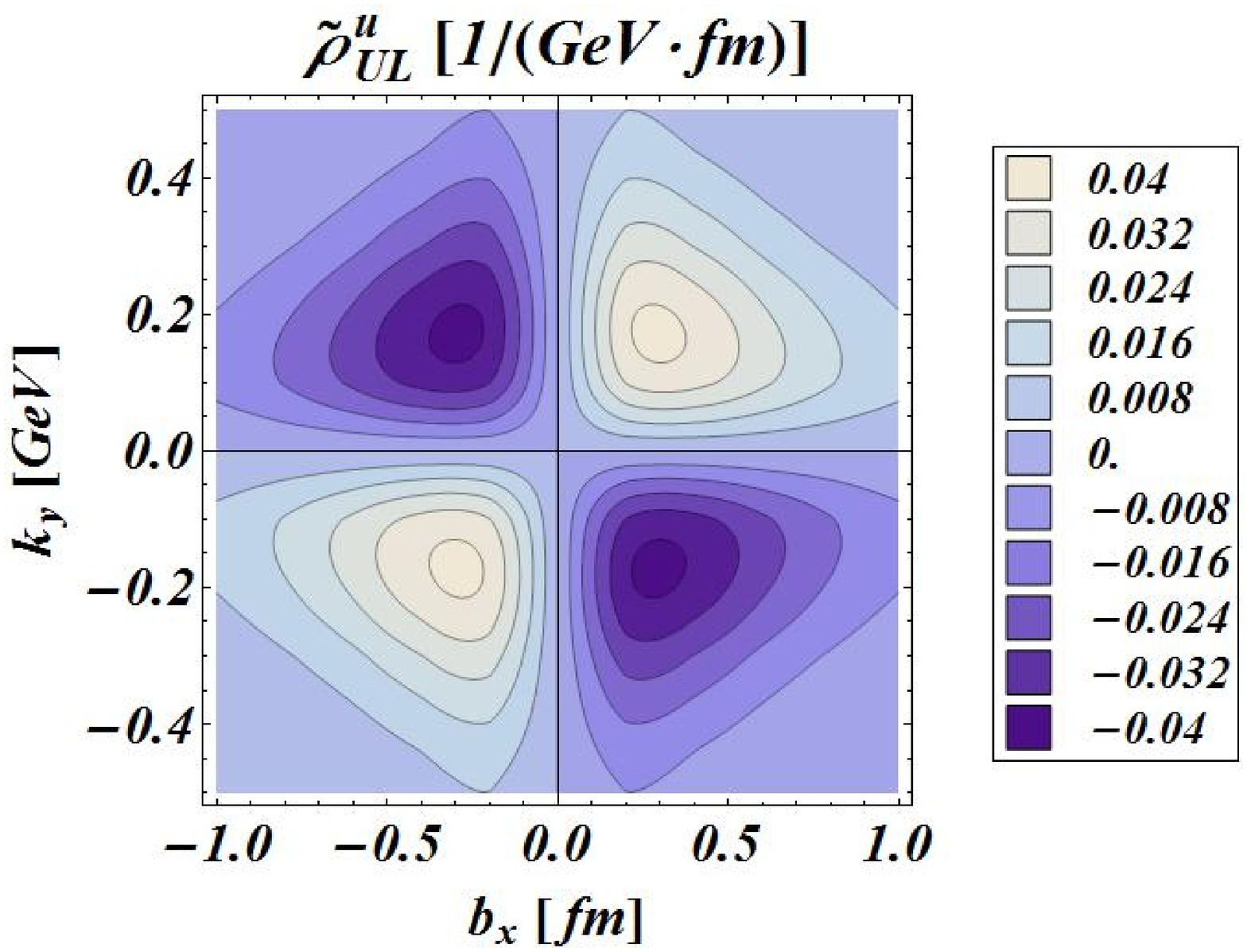}
		\includegraphics[width=.49\textwidth]{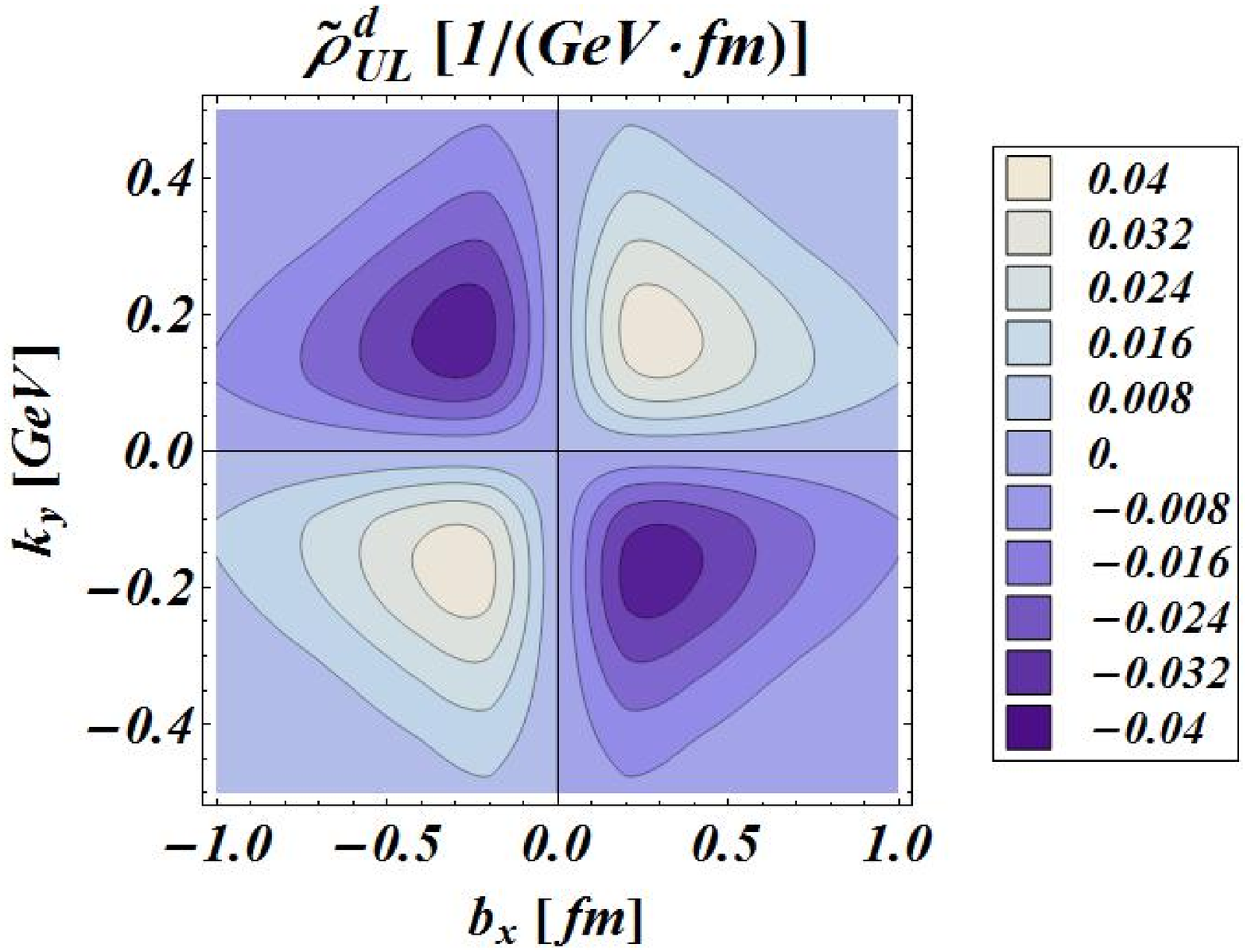}
		\caption{\footnotesize{The distortions of the mixed transverse densities $\tilde\rho(b_x,k_y)$ due to the spin (pointing out of the plane) of $u$ quarks (left panel) and $d$ quarks (right panel) in an unpolarized proton.}}
		\label{fig7}
\end{figure}

We now discuss $\rho_{UL}(\vec b_\perp,\vec k_\perp)$, the distortion of the transverse Wigner distribution due to the longitudinal polarization of quarks in an unpolarized proton. In Fig.~\ref{fig6} we show the distortions, both in impact-parameter space with fixed transverse momentum $\vec k_\perp=k_\perp\,\hat e_y$ and $k_\perp=0.3$ GeV (upper panels) and in transverse-momentum space with fixed impact parameter $\vec b_\perp=b_\perp\,\hat e_y$ and $b_\perp=0.4$ fm (lower panels), of the distributions of $u$ (left panels) and $d$ quarks (right panels). The corresponding distortions of the mixed transverse densities $\tilde\rho(b_x,k_y)$ are shown in Fig.~\ref{fig7}, for $u$ (left panel) and for $d$ (right panel) quarks.

Like in the case of the $\rho_{LU}$ distributions, the dipole and quadrupole structures are due to the explicit factor $\epsilon^{ij}_\perp k^i_\perp\tfrac{\partial}{\partial b_\perp^j}$ in Eq.~\eqref{UL}. We learn from these figures that the quark OAM and the quark spin tend to be aligned for both $u$ and $d$ quarks. The size of the distributions are similar for $u$ and $d$ quarks. However, since there are effectively twice more $u$ quarks than $d$ quarks in the proton, the alignment is more pronounced for the $d$ quarks than the $u$ quarks. The correlation  $C^q_z$ between quark spin and OAM in the $\hat e_z$ direction can be calculated using the definition in Eq.~\eqref{corrSpinOAM}. The results for both the LCCQM and the $\chi$QSM are given in Table \ref{OAMcorrtable}. As anticipated, we  find $C^u_z>0 $  and $C^d_z>0$, with  larger values in the LCCQM than the $\chi$QSM.
\begin{table}[th!]
\begin{center}
\caption{\footnotesize{The results for quark spin-OAM correlation $C^q_z$ (see Eq.~\eqref{corrSpinOAM}) and anomalous tensor magnetic moment $\kappa^q_T$ obtained in the LCCQM and the $\chi$QSM for $u$-, $d$- and total ($u+d$) quark contributions.}}\label{OAMcorrtable}
\begin{tabular}{@{\quad}c@{\quad}c@{\quad}|@{\quad}c@{\quad}c@{\quad}c@{\quad}|@{\quad}c@{\quad}c@{\quad}c@{\quad}}\whline
\multicolumn{2}{@{\quad}c@{\quad}|@{\quad}}{Model}&\multicolumn{3}{c@{\quad}|@{\quad}}{LCCQM}&\multicolumn{3}{c@{\quad}}{$\chi$QSM}\\
\multicolumn{2}{@{\quad}c@{\quad}|@{\quad}}{$q$}&$u$&$d$&Total&$u$&$d$&Total\\
\hline
$C^q_z$&Eq.~\eqref{corrSpinOAM}&$0.227$&$0.187$&$0.414$&$0.130$&$0.109$&$0.239$
\\\hline
$\kappa^q_T$&&$3.947$&$2.581$&$6.528$&$3.832$&$2.582$&$6.414$\\
\whline
\end{tabular}
\end{center}
\end{table}

In Table~\ref{OAMcorrtable} we also give the results for the tensor anomalous magnetic moment $\kappa^q_T$ which measures the correlation between the transverse spin and the transverse OAM of the quark in an unpolarized nucleon, as observed in the IPDs for  transversely polarized quarks in an unpolarized nucleon~\cite{Burkardt:2005hp,Pasquini:2007xz}. We find $\kappa_T^u > \kappa^d_T>0$, which coincides with the pattern of $C^q_z$. However, at variance with  $C^q_z$, the values for $\kappa_T^q$ from the LCCQM~\cite{Pasquini:2005dk} and the $\chi$QSM~\cite{Lorce:2011dv} are very similar.

\subsection{Longitudinally Polarized Quarks in a Longitudinally Polarized Nucleon}
\label{section-3d}

\begin{figure}[t!]
	\centering
		\includegraphics[width=.49\textwidth]{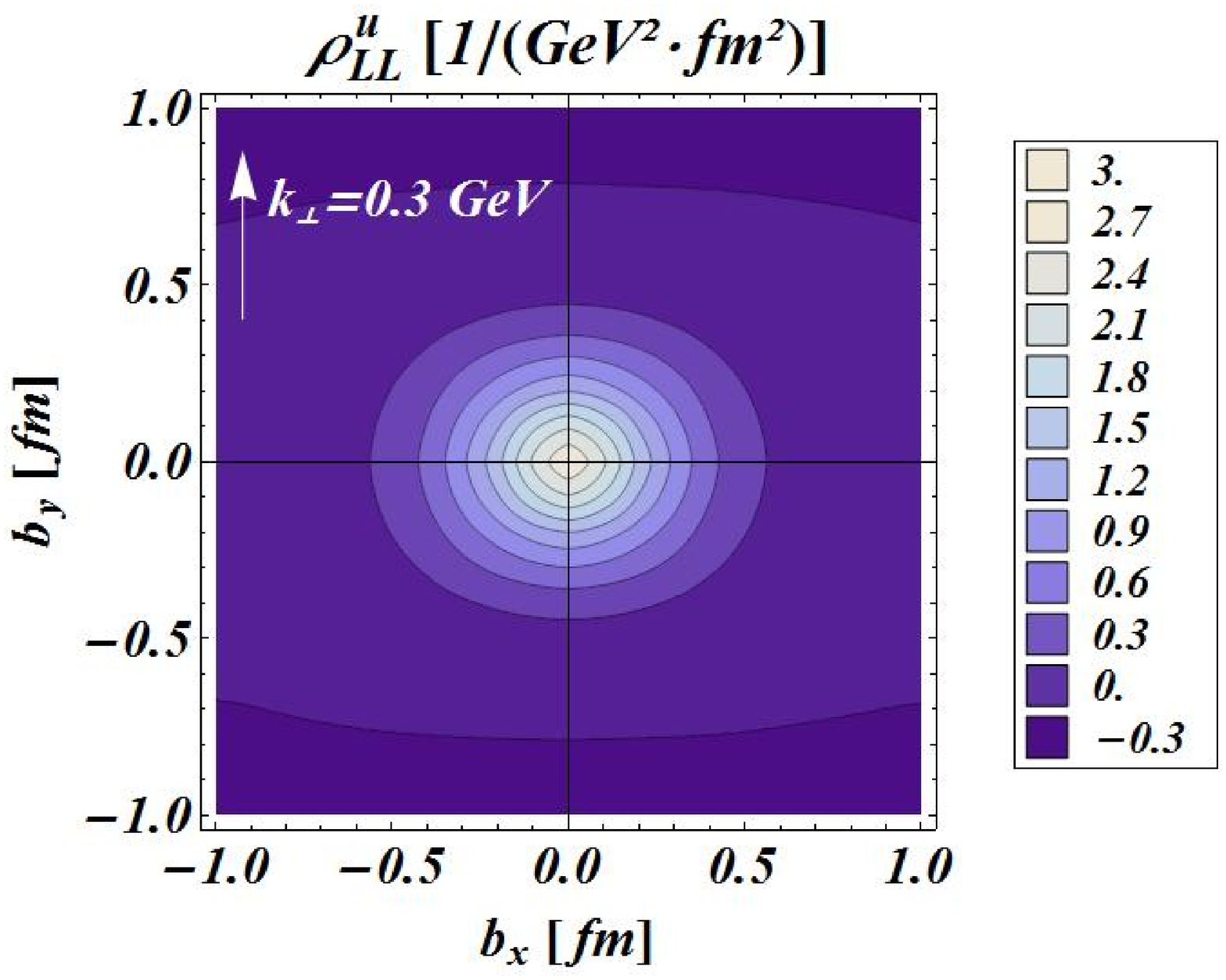}
		\includegraphics[width=.49\textwidth]{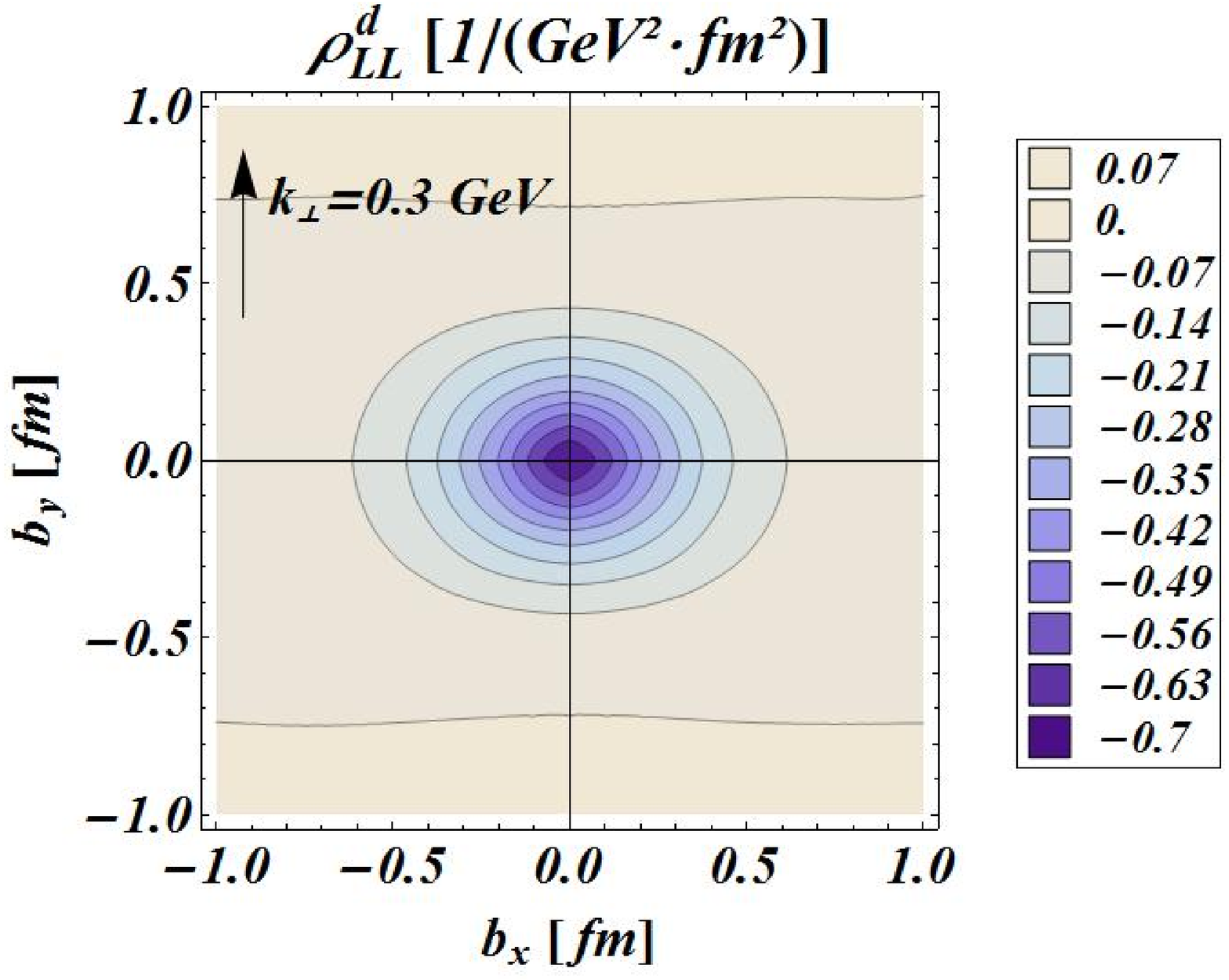}
		\includegraphics[width=.49\textwidth]{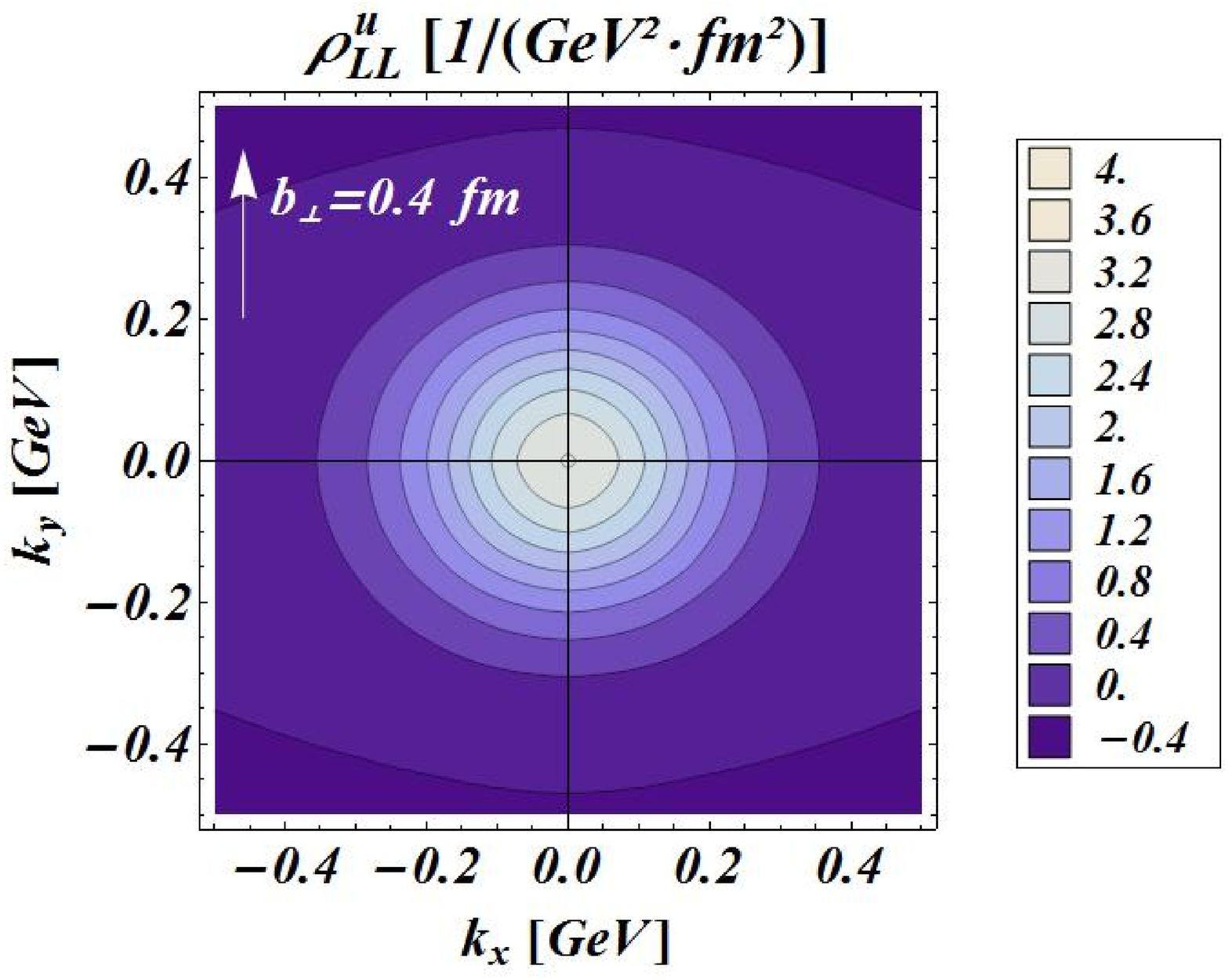}
		\includegraphics[width=.49\textwidth]{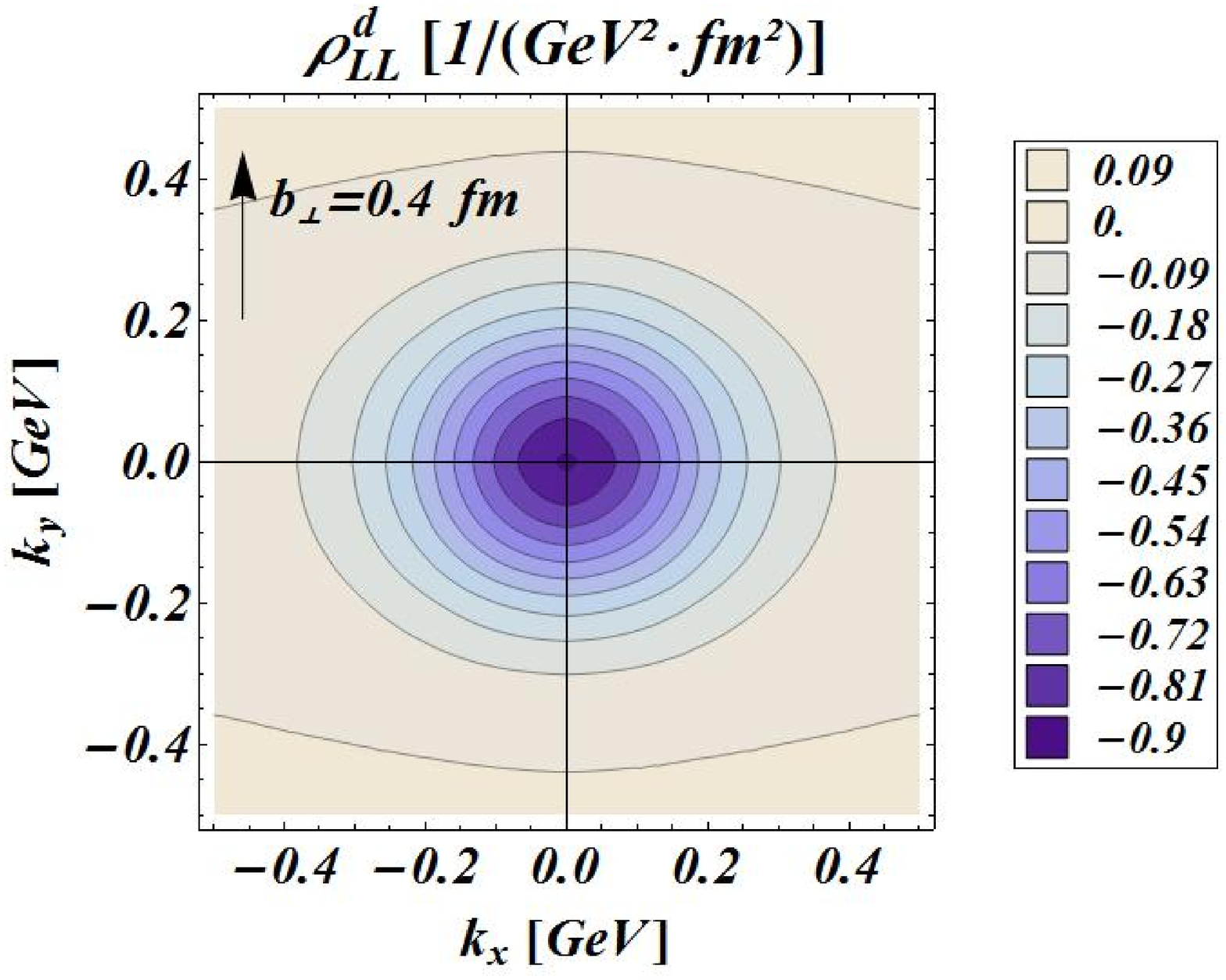}
\caption{\footnotesize{The distortions of the transverse Wigner distributions due to the correlation between the quark spins and the proton spin (pointing out of the plane). Upper panels: distortions in impact-parameter space with fixed transverse momentum $\vec k_\perp=k_\perp\,\hat e_y$ and $k_\perp=0.3$ GeV. Lower panels: distortions in transverse-momentum space with fixed impact parameter $\vec b_\perp=b_\perp\,\hat e_y$ and $b_\perp=0.4$ fm. The left (right) panels show the results for $u$ ($d$) quarks.}}
		\label{fig8}
\end{figure}

\begin{figure}[th!]
	\centering
		\includegraphics[width=.49\textwidth]{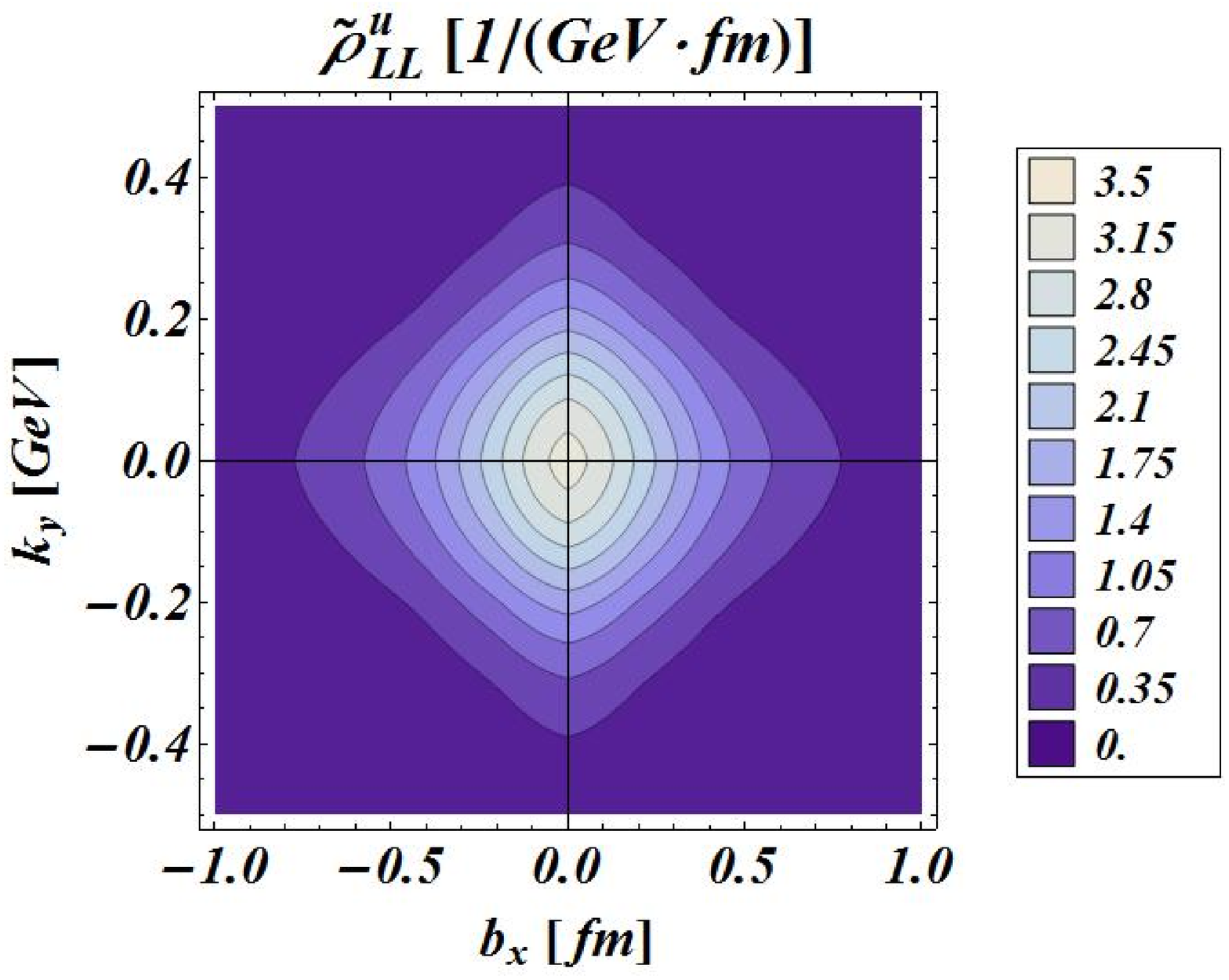}
		\includegraphics[width=.49\textwidth]{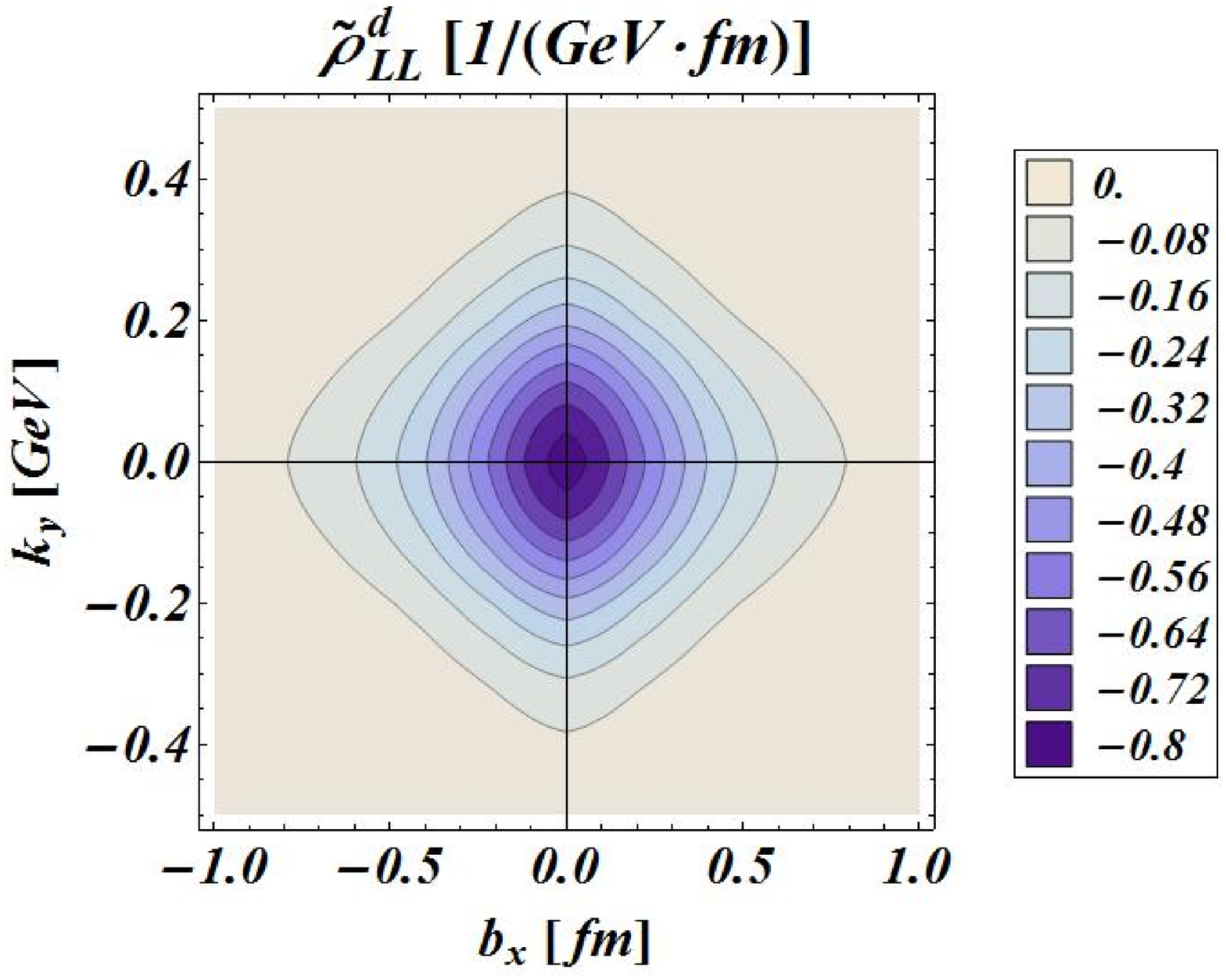}
		\caption{\footnotesize{The distortions of the mixed transverse densities $\tilde\rho(b_x,k_y)$ due to the correlation between the spin of $u$ quarks (left panel) and $d$ quarks (right panel), and the proton spin (pointing out of the plane).}}
		\label{fig9}
\end{figure}

We proceed with the discussion of $\rho_{LL}(\vec b_\perp,\vec k_\perp)$, the distortion of the transverse Wigner distribution due to the correlation between the longitudinal polarizations of the quarks and the proton. In Fig.~\ref{fig8} we show the distortions, both in impact-parameter space with fixed transverse momentum $\vec k_\perp=k_\perp\,\hat e_y$ and $k_\perp=0.3$ GeV  (upper panels) and in transverse-momentum space with fixed impact parameter $\vec b_\perp=b_\perp\,\hat e_y$ and $b_\perp=0.4$ fm (lower panels), of the distributions of $u$ (left panels) and $d$ quarks (right panels). 
The corresponding distortions of the mixed transverse densities $\tilde\rho(b_x,k_y)$ are shown in Fig.~\ref{fig9}.
As one already knows from the axial charges, the $u$-quark polarization tends to be parallel to the nucleon spin, while the $d$-quark polarization tends to be antiparallel. Accordingly, the distributions are positive for $u$ quarks and negative for $d$ quarks. The new information is about the distribution in phase space of these polarizations (see Eq.~\eqref{intLL}). It appears that the quark polarization receives its main contribution from the central region of the phase space. Interestingly, the average quark polarization changes sign for sufficiently large $b_\perp$ or $k_\perp$, preferably when $\vec b_\perp$ and $\vec k_\perp$ are aligned (see Fig.~\ref{fig8}). From Eq.~\eqref{LL} we see that $\rho_{LL}(\vec b_\perp,\vec k_\perp)=\rho_{LL}(b_\perp,k_\perp,\vec k_\perp\cdot\vec b_\perp)$, explaining the left-right symmetry in Fig~\ref{fig8}. It follows that $\rho_{LL}$ cannot contribute to the net quark OAM, as required by the isotropy of space (see Eq.~\eqref{OAMLL}).

\begin{figure}[t!]
	\centering
		\includegraphics[width=.49\textwidth]{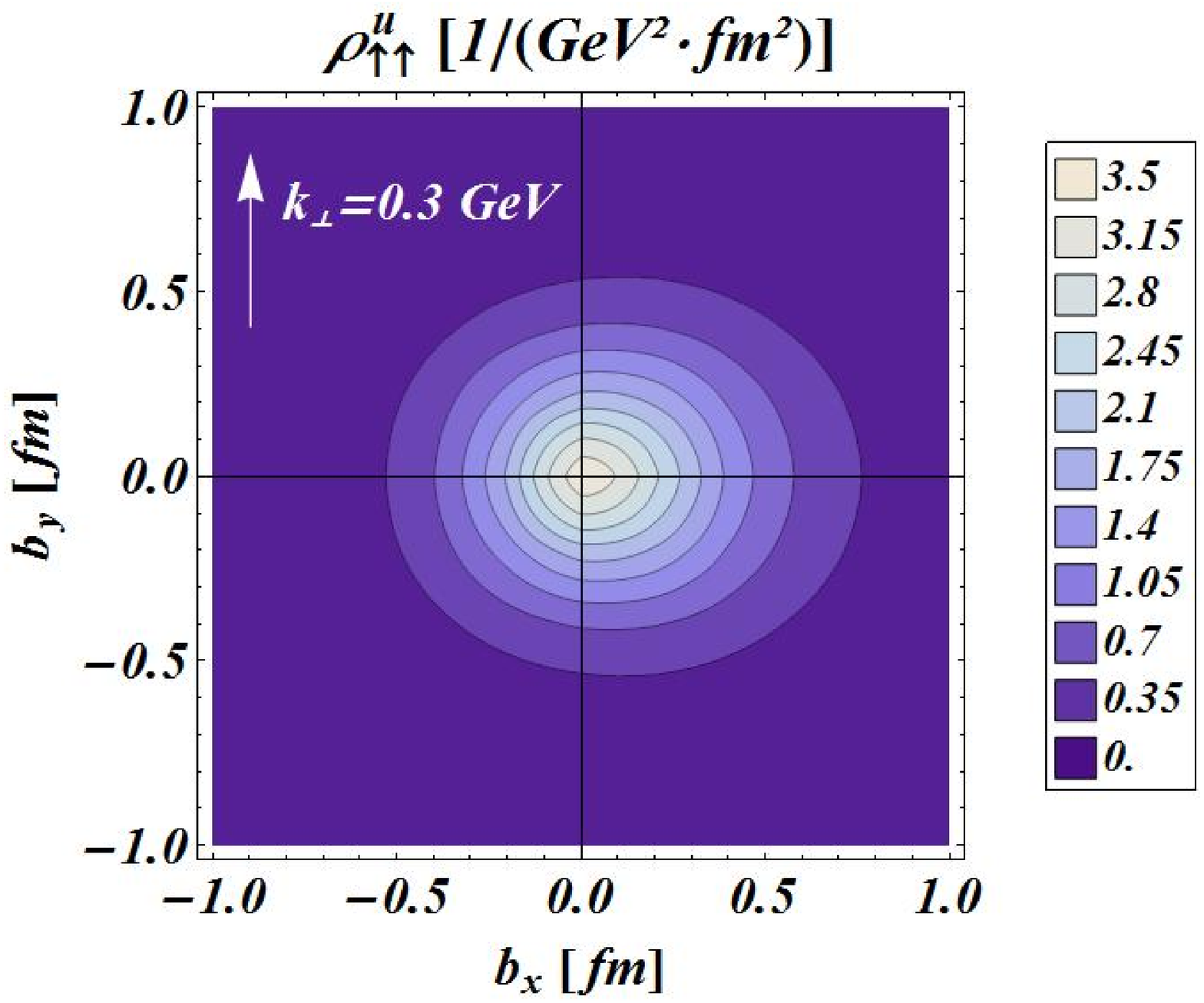}
		\includegraphics[width=.49\textwidth]{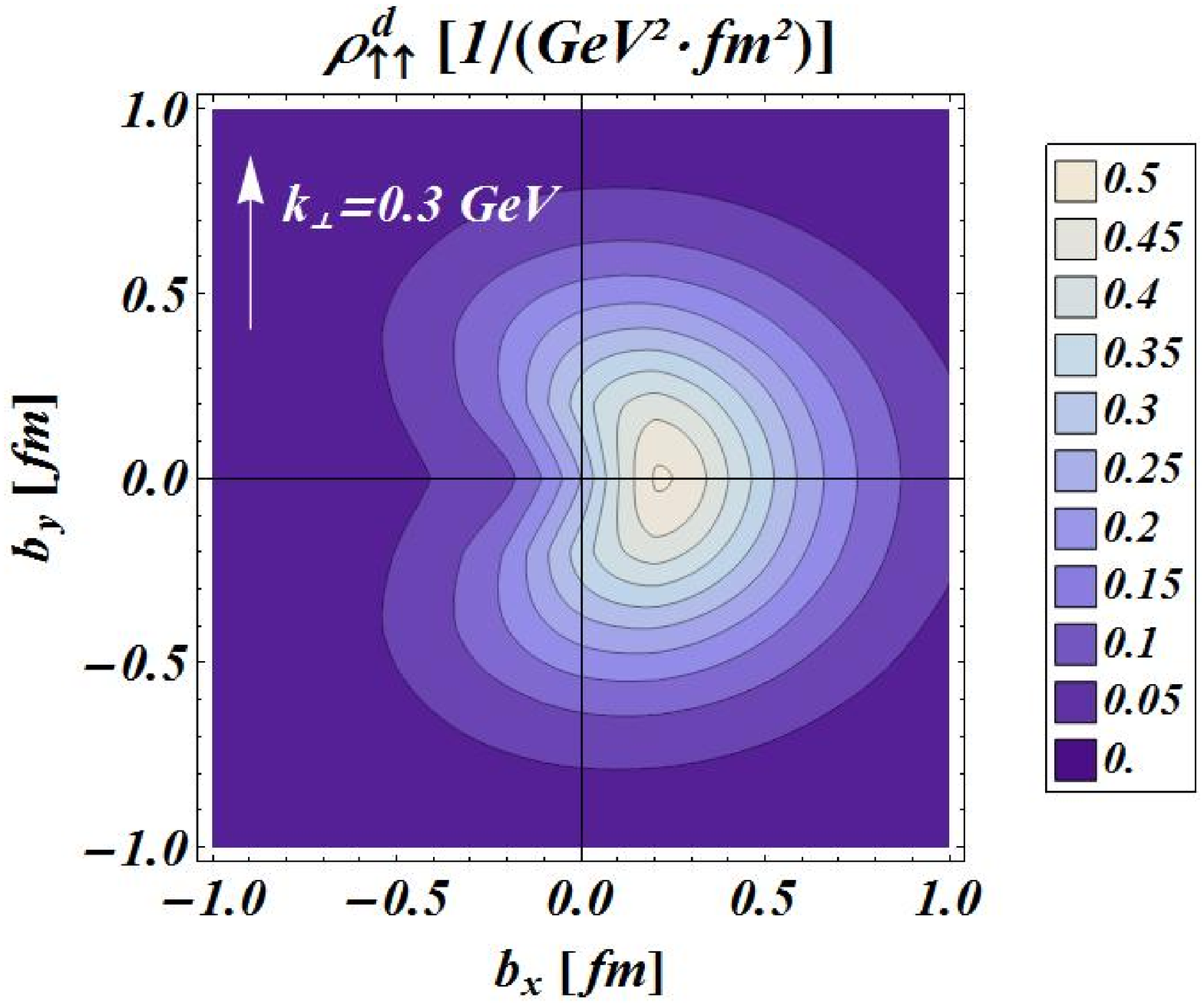}
		\includegraphics[width=.49\textwidth]{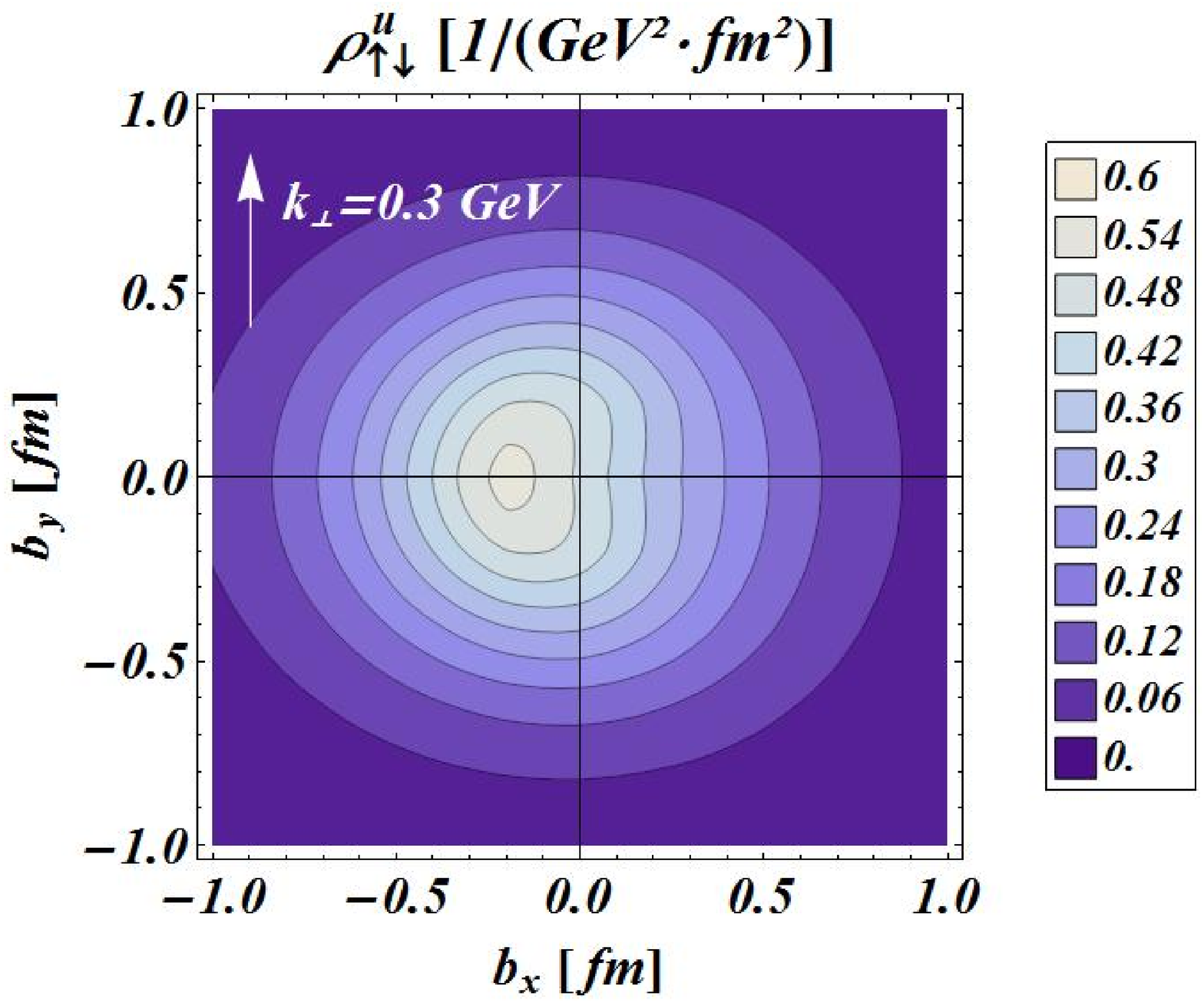}
		\includegraphics[width=.49\textwidth]{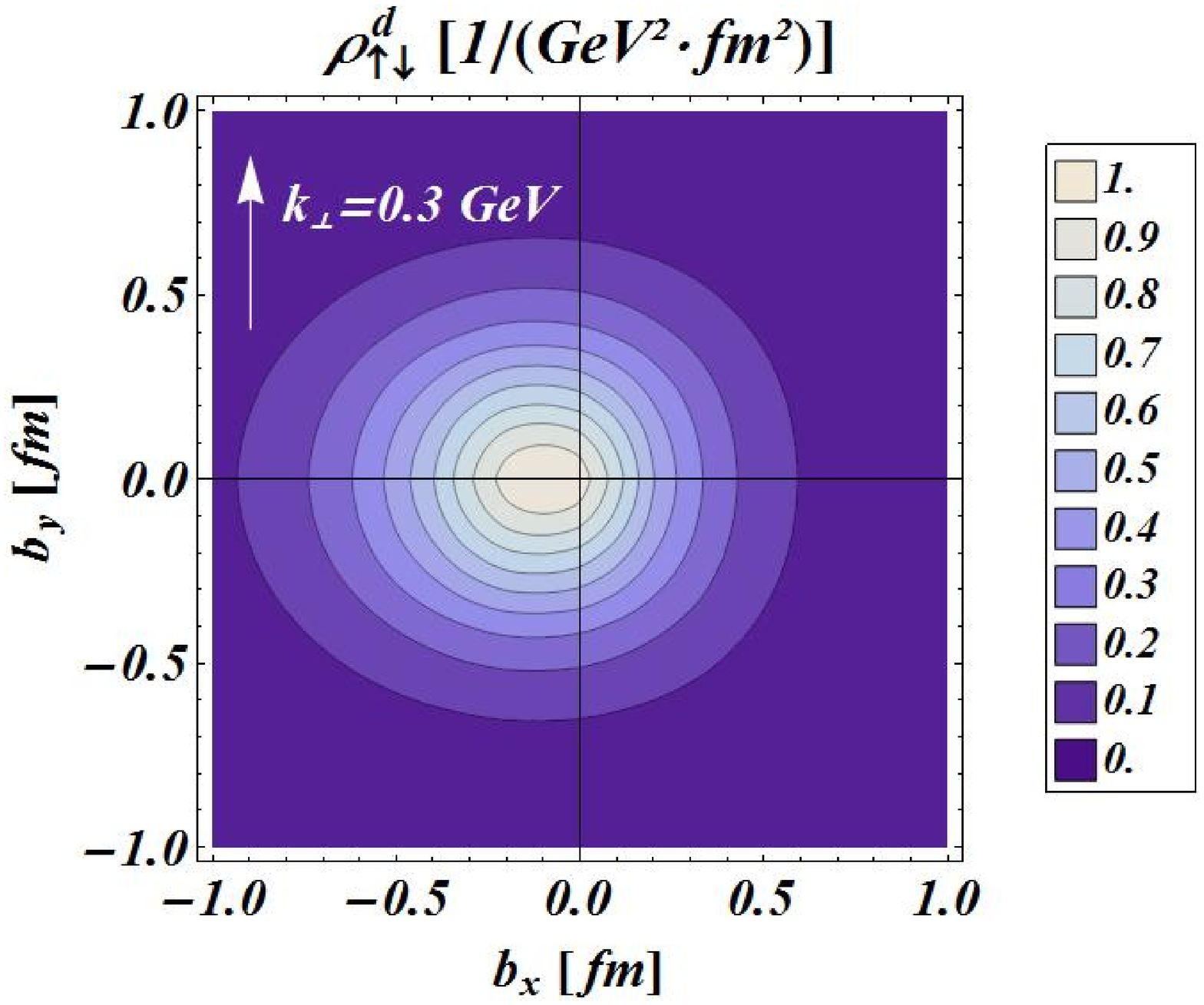}
\caption{\footnotesize{The transverse Wigner distributions of longitudinally polarized quarks in a longitudinally polarized proton ($\Lambda=\uparrow$ pointing out of the plane) in impact-parameter space with fixed transverse momentum $\vec k_\perp=k_\perp\,\hat e_y$ and $k_\perp=0.3$ GeV. Upper panels: distributions of quarks with polarization parallel to the nucleon spin ($\lambda=\uparrow$). Lower panels: distributions of quarks with polarization antiparallel to the nucleon spin ($\lambda=\downarrow$). The left (right) panels show the results for $u$ ($d$) quarks.}}
		\label{fig10}
\end{figure}
\begin{figure}[t!]
	\centering
		\includegraphics[width=.49\textwidth]{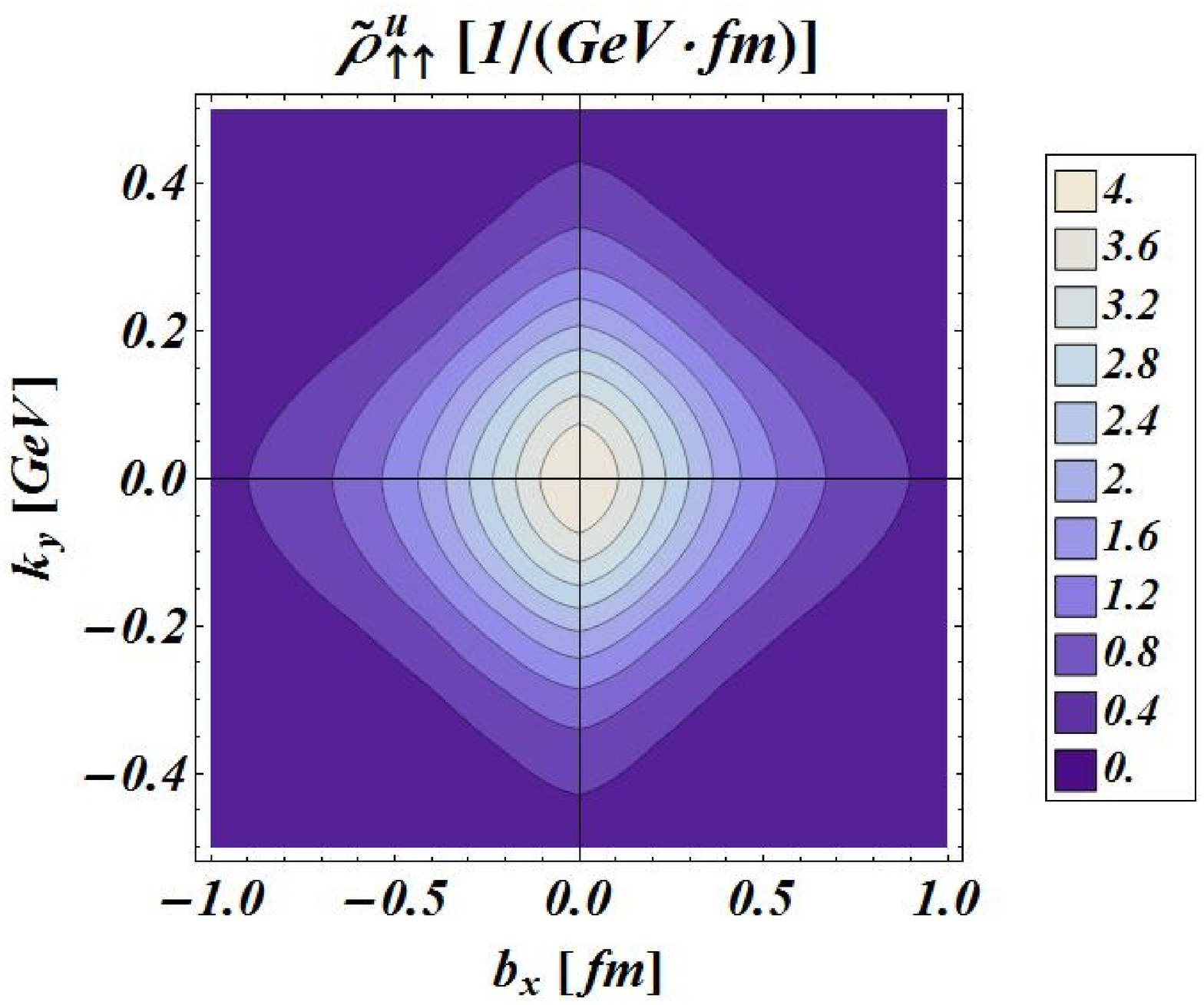}
		\includegraphics[width=.49\textwidth]{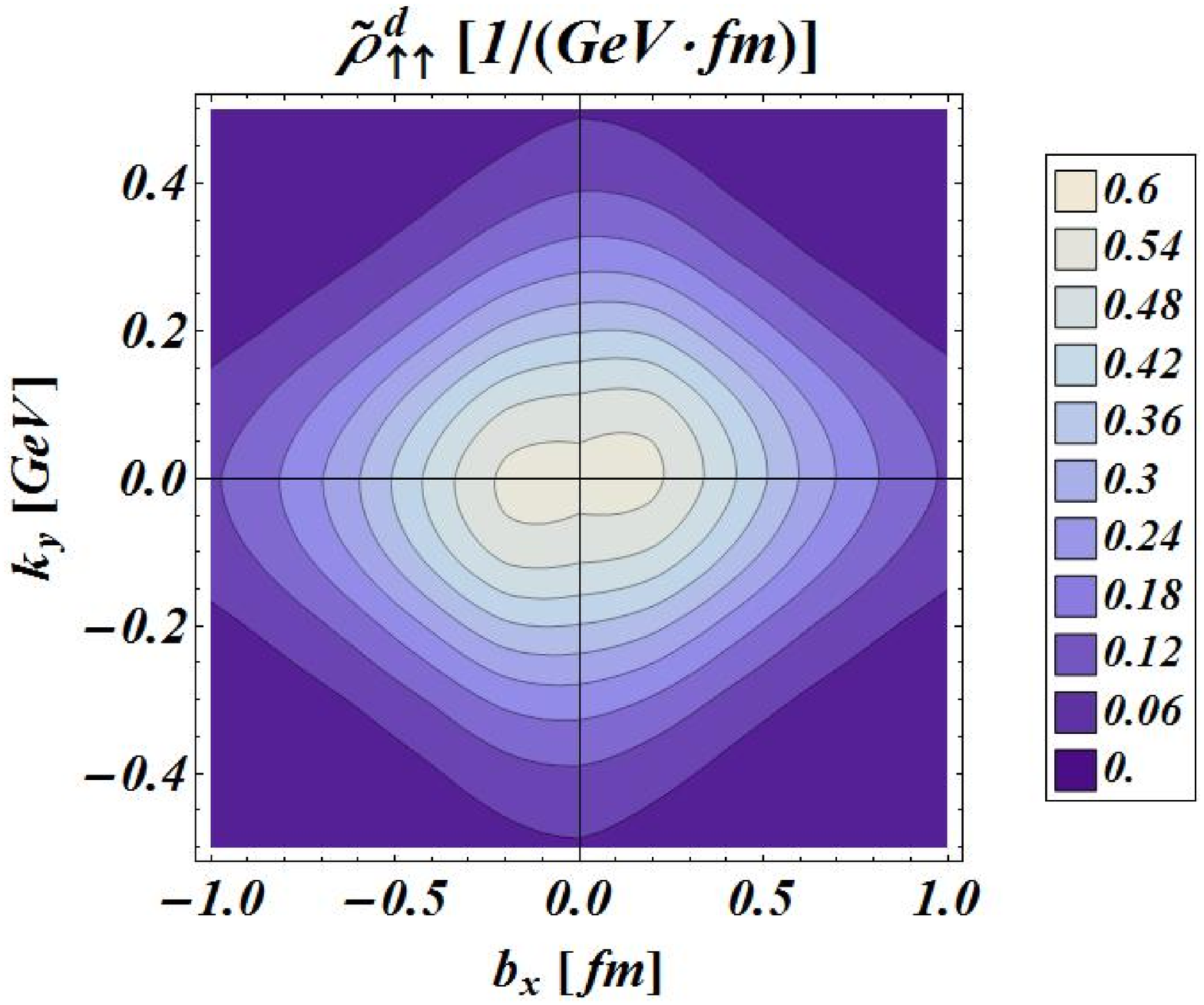}
		\includegraphics[width=.49\textwidth]{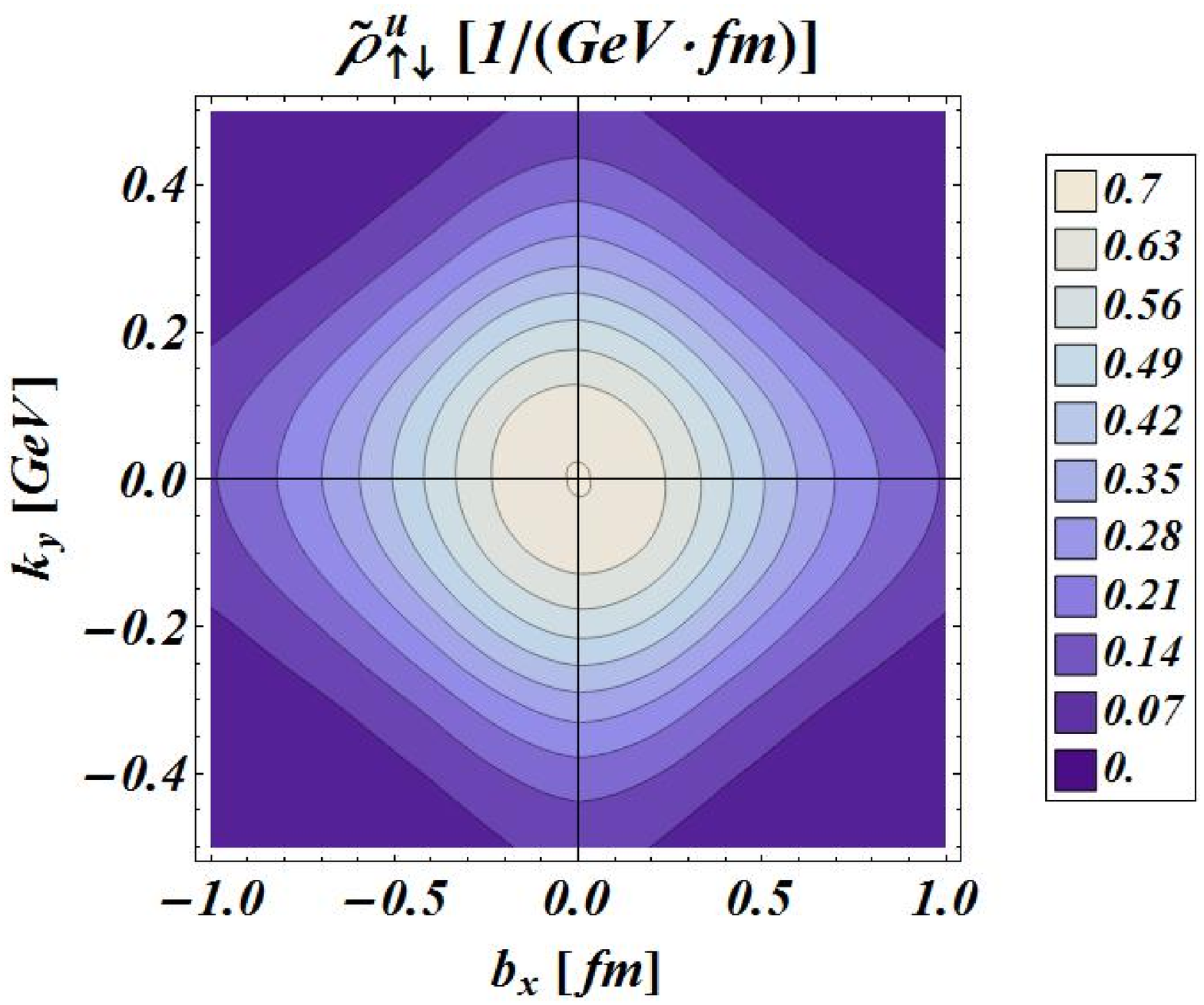}
		\includegraphics[width=.49\textwidth]{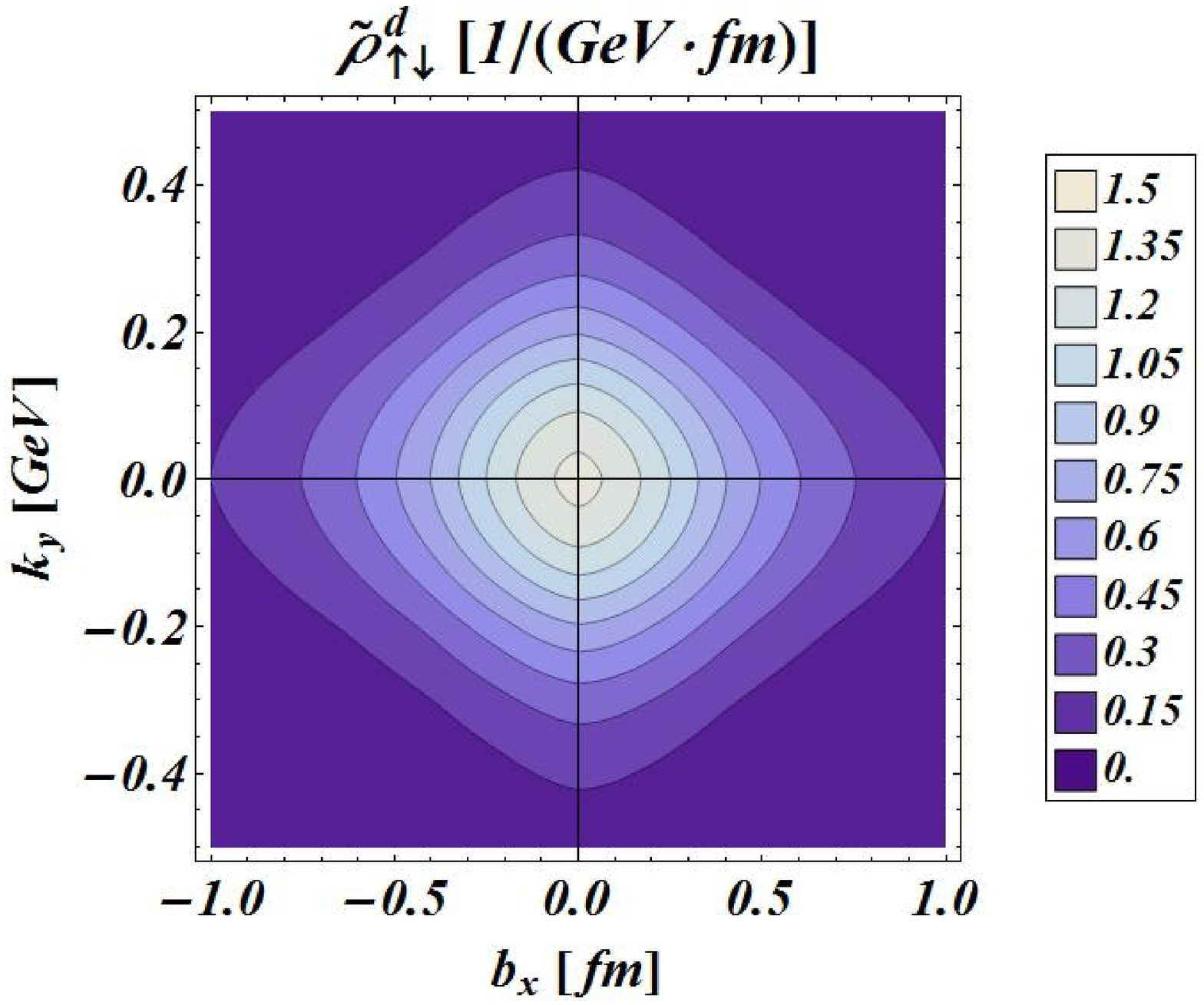}
\caption{\footnotesize{The mixed transverse densities $\tilde\rho_{\Lambda\lambda}(b_x,k_y)$ of longitudinally polarized quarks in a longitudinally polarized proton ($\Lambda=\uparrow$ pointing out of the plane). Upper panels: probability densities of quarks with polarization parallel to the nucleon spin ($\lambda=\uparrow$). Lower panels: probability densities of quarks with polarization antiparallel to the nucleon spin ($\lambda=\downarrow$). The left (right) panels show the results for $u$ ($d$) quarks.}}
		\label{fig11}
\end{figure}

Combining as in Eq.~\eqref{WD} the Wigner distribution of unpolarized quarks in an unpolarized proton $\rho_{UU}$ with the distortions $\rho_{LU}$, $\rho_{UL}$ and $\rho_{LL}$, we obtain the Wigner distribution $\rho_{\Lambda\lambda}$ of longitudinally polarized quarks in a longitudinally polarized proton. In  Fig.~\ref{fig10}, the transverse Wigner distributions of $u$ and $d$ quarks with polarization $\lambda=\uparrow,\downarrow$ in a proton with polarization $\Lambda=\uparrow$ are shown in the impact-parameter space with fixed transverse momentum $\vec k_\perp=k_\perp\,\hat e_y$ and $k_\perp=0.3$ GeV. The corresponding mixed transverse densities $\tilde\rho_{\Lambda\lambda}(b_x,k_y)$ are shown in Fig.~\ref{fig11}.
The deformation induced by the quark and nucleon polarizations is clearly visible in the sideway shifts of the distributions in Fig~\ref{fig10}. In particular, when the quark and nucleon polarizations are parallel (antiparallel) the shift is in the positive (negative) $\hat b_x$ direction, see upper (lower) panels. 

We learned from $\rho_{LU}$ and $\rho_{UL}$ that the $u$-quark OAM tends to be aligned with both the quark and proton polarizations. When the $u$ quark has polarization parallel to the nucleon spin, the contributions $\rho_{LU}$ and $\rho_{UL}$ interfere constructively resulting in a sideway shift in the positive $\hat b_x$ direction. When the $u$ quark has polarization antiparallel to the nucleon spin, the contributions $\rho_{LU}$ and $\rho_{UL}$ interfere destructively. Since the correlation between the OAM and the quark spin is stronger than the correlation between the OAM and the nucleon spin (see Figs.~\ref{fig4} and \ref{fig6}), it results a sideway shift in the negative $\hat b_x$ direction. For the $d$ quark, we learned from $\rho_{LU}$ and $\rho_{UL}$ that the OAM tends to be aligned with the quark polarization but antialigned with the nucleon polarization. When the $d$ quark has polarization parallel to the nucleon spin, the contributions $\rho_{LU}$ and $\rho_{UL}$ interfere destructively. Once again, we can see from Figs.~\ref{fig4} and \ref{fig6} that the correlation between the OAM and the quark spin is stronger than the correlation between the OAM and the nucleon spin, resulting in a sideway shift in the positive $\hat b_x$ direction. When the $d$ quark has polarization antiparallel to the nucleon spin, the contributions $\rho_{LU}$ and $\rho_{UL}$ interfere constructively resulting in a sideway shift in the negative $\hat b_x$ direction. The sideway shifts are more apparent in $\rho^u_{\uparrow\downarrow}$ and $\rho^d_{\uparrow\uparrow}$ because the contributions $\rho_{UU}$ and $\rho_{LL}$ partially cancel in these cases. In an analogous way we can understand the densities in the $(b_x,k_y)$ plane shown in Fig.~\ref{fig11}.

\section{Conclusions}
\label{section-4}

In this work we presented a study of the quark Wigner functions which provide the full phase-space description of the quark distributions in the nucleon. Using the light-front formalism, we derived the Wigner distributions as two-dimensional Fourier transforms of the GTMDs from the transverse-momentum transfer $\vec\Delta_\perp$ to the impact parameter $\vec b_\perp$. Therefore these distributions provide us with images of the nucleon in five dimensions, namely two position and three momentum coordinates. This derivation is not spoiled by relativistic corrections and is completely analogous to the interpretation of transverse charge densities and impact-parameter dependent parton distributions as two-dimensional Fourier transforms of form factors and GPDs, respectively. However, Wigner distributions can not have a strict probabilistic interpretation, since Heisenberg uncertainty relations forbid to localize a particle and to determine its momentum at the same time. Accordingly, the Wigner distributions are not positive definite. Only in particular limits one can  recover a density interpretation. This is the case for the known projections of the Wigner distributions to the three-dimensional densities in the momentum or in the impact-parameter space at fixed $x$, corresponding to TMDs and IPDs, respectively. On top of them, we introduced two new types of densities mapping the nucleon as functions of one spatial and one momentum variables in the transverse plane which are not conjugated, and therefore not constrained by the uncertainty principle.

In general the GTMDs are complex-valued functions. However their two-dimensional Fourier transforms are always real-valued functions, in accordance with the interpretation of the Wigner distributions as phase-space distributions. At leading twist, there are 16 Wigner distributions which can be disentangled by varying the nucleon and quark polarizations. We focused on the four cases without transverse polarizations, namely the distributions of unpolarized/longitudinally polarized quarks in an unpolarized/longitudinally polarized nucleon. Furthermore, we considered only the quark contribution, neglecting all the gauge-field degrees of freedom. In this case, the imaginary part of the GTMDs is zero, and by hermiticity and time-reversal invariance the corresponding Wigner distributions are even functions of $\vec k_\perp\cdot\vec b_\perp$. 

The results for the Wigner distributions within a light-cone constituent quark model and the light-cone version of the chiral quark-soliton model are very similar, and allowed us to sketch some general features about the behavior of the quarks in the nucleon when observed in the $\vec b_\perp$ plane at fixed $\vec k_\perp$, or in the $\vec k_\perp$ plane at fixed $\vec b_\perp$. In particular, the Wigner distributions of unpolarized quarks in an unpolarized nucleon are not axially symmetric. This deformation can be explained with naive semi-classical arguments, as a consequence of confinement which limits the radial motion of the quark with respect to its orbital motion. The effect becomes more pronounced at larger values of $k_\perp$ ($b_\perp$) in the impact-parameter (momentum) space, as found by calculating the corresponding quadrupole distortions. Furthermore, we observed that the spread of the distributions is smaller for $u$ quarks than for $d$ quarks, especially in the $\vec b_\perp$ space, revealing that the $u$ quarks are more concentrated at the center of the proton, while the $d$-quark distribution has a tail which extends further at the periphery of the proton.

The case of unpolarized quarks in a longitudinally polarized nucleon is particularly interesting because it allows us to calculate the phase-space average of the quark OAM. The corresponding values within the LCCQM and the light-cone $\chi$QSM have been compared with the results for the OAM from the Ji's sum rule, and from the definition in terms of the $h_{1T}^\perp$ TMD. We found that in models without gauge fields these three definitions give the same values for the total quark contribution to the OAM, while they differ for the individual $u$ and $d$ quark-flavor contributions. A peculiar result of the light-cone $\chi$QSM is that the isovector combination ($u-d$) of the OAM calculated from the Ji's sum rule is found to be negative, in agreement with lattice calculations and at variance with most of quark models.

The distortion due to the longitudinal polarization of quarks in an unpolarized nucleon allowed us to study the correlation between the quark spin and OAM. This correlation, taking into account the effective number of $u$ and $d$ quarks, has been found to be stronger for $d$ quarks than for $u$ quarks. The same behavior has also been observed for the values of the tensor anomalous magnetic moments which measure the correlation between the transverse spin and the transverse OAM of quarks in an unpolarized nucleon.

In the case of the distortion due to the correlation between the quark and nucleon spins, we were able to study the distributions of the axial charge in the phase space. They are positive for $u$ quarks and negative for $d$ quarks, and receive the main contribution from the central region of the phase space. An interesting finding is the fact that the average  polarization of $d$ quarks changes sign for sufficiently large $b_\perp$ and $k_\perp$, preferably when $\vec b_\perp$ and $\vec k_\perp$ are aligned.

Finally, taking into account all the four contributions discussed above, we visualized the combined effects induced on the distributions by the longitudinal polarizations of the quarks and nucleon. In all the examples we have studied, the prominent role of OAM and its correlation with quark and nucleon polarizations in shaping the quark distributions in the phase space has clearly emerged. Besides specific features related to the quark models we used for discussing the results, we tried to emphasize the physical content of the Wigner distributions, and in particular the new information encoded in these distributions. Further studies of the Wigner distributions in different theoretical models can provide new insights on the quark and gluon dynamics. On the other hand, although Wigner distributions are not directly measurable from experiments, any new information on GPDs and TMDs extracted from experiments can be used to further constrain the multidimensional image of the nucleon in the quantum phase space.

\section*{Acknowledgments}

C. L. is thankful to INFN and the Department of Nuclear and Theoretical Physics of the University of Pavia for their hospitality. This work was supported in part by the Research Infrastructure Integrating Activity ``Study of Strongly Interacting Matter'' (acronym HadronPhysics2, Grant Agreement n. 227431) under the Seventh Framework Programme of the European Community, by the Italian MIUR through the PRIN 2008EKLACK ``Structure of the nucleon: transverse momentum, transverse spin and orbital angular momentum''.

\end{document}